\documentclass[trackchanges,twocolumn]{aastex}

\let\tablenum\relax

\usepackage{natbib,epsfig,siunitx,url,graphicx,amsmath,footnote,float,xcolor,cases,lineno}
\begin{document}

\submitted{Accepted for publication in \textit{AJ}}

\title{Stability of Neptune's distant resonances in the presence of Planet Nine}

\author{Matthew S. Clement\altaffilmark{1}, \& Scott S. Sheppard\altaffilmark{1}}

\altaffiltext{1}{Earth and Planets Laboratory, Carnegie Institution
for Science, 5241 Broad Branch Road, NW, Washington, DC
20015, USA}
\altaffiltext{2}{HL Dodge Department of Physics Astronomy, University of Oklahoma, Norman, OK 73019, USA}
\altaffiltext{*}{corresponding author email: mclement@carnegiescience.edu}

\setcounter{footnote}{0}
\begin{abstract}

Trans-Neptunian Objects (TNOs) in the scattered disk with 50 $\lesssim a \lesssim$ 100 au are thought to cluster near Neptune's n:1 resonances (e.g: 3:1, 4:1, and so on).  While these objects spend lengthy periods of time at large heliocentric distances, if their perihelia remain less than around 40 au, their dynamical evolution is still largely coupled to Neptune's.  Conversely, around a dozen extreme TNOs with $a \gtrsim$ 250 au and detached perihelia seem to exist in a regime where they are too distant to be affected by the giant planets, and too close for their dynamics to be governed by external forces.  Recent work suggests that the apparent alignment of these orbits in physical space is a signature of gravitational shepherding by a distant massive planet.  In this paper, we investigate the evolution of TNOs in each of Neptune's n:1 resonances between the 3:1 and 14:1.  We conclude that both resonant and non-resonant objects beyond the 12:1 near $\sim$157 au are removed rather efficiently via perturbations from the hypothetical Planet Nine.  Additionally, we uncover a population of simulated TNOs with $a\lesssim$ 100 au, 40 $\lesssim$$q$$\lesssim$ 45 au and low inclinations that experience episodes of resonant interactions with both Neptune and Planet Nine.  Finally, we simulate the evolution of observed objects with $a>$ 100 au and identify several TNOs that are potentially locked in n:1 resonances with Neptune; including the most distant known resonant candidates 2014 JW$_{80}$ and 2014 OS$_{394}$ that appear to be in the 10:1 and 11:1 resonances, respectively.  Our results suggest that the detection of similar remote objects might provide a useful constraint on hypotheses invoking the existence of additional distant planets.
\end{abstract}

\section{Introduction}

It is well established that the gravitational interplay between Neptune and diminutive objects in the Kuiper Belt resulted in the planet's net outward migration in the time since the solar system's birth \citep{fer84,thommes99,hahn99}.  Through this process trans-Neptunian objects (TNOs) fill and populate the dominant mean motion resonances (MMRs) with Neptune \citep{malhotra93,malhotra95}; thereby reshaping the primordial Kuiper Belt into the structure of orbits observed today \citep[e.g.:][]{nesvorny15a,nesvorny16_grainy,kaib16}.  Thus, a record of the solar system's earliest dynamics is somewhat fossilized in the Kuiper Belt's complex distribution of orbits, and detailed observations \citep[e.g.:][]{sedna,sheppard16,bannister16_ossos} of the objects in the region serve as a sounding board for the verification of proposed solar system evolutionary hypotheses \citep{levison08,nesvorny15b,kaib19}.

While the innermost regions of the Kuiper Belt are beginning to be constrained observationally, the population of detected objects with $a \gtrsim$ 100 au and perihelia beyond Neptune remains small (86 versus 3,018 objects with 30 $<a<$ 100 au and $q>$ 30 au according to the Minor Planet Center database\footnote{Minor Planet Center (MPC) queried on 28 December 2020 throughout this manuscript}).  Consequentially, the precise resonant structure \citep[e.g.:][]{chiang03,gladman12,holman18,volk18_9_1} of the distant Kuiper Belt remains largely theoretical, and is thus inferred via numerical simulations of Neptune's orbital migration \citep{kaib16,nesvorny16_50au} before and after the giant planet instability \citep{nesvorny12,deienno17,clement20_instb}.  Investigations following in this mold typically find that distant TNOs cluster about the successive n:1 MMRs with Neptune \citep[3:1, 4:1, and so on:][]{kaib16} through a process known as ``resonance sticking'' \citep{gallardo06,lykawka07}.  The relative densities of objects in or near these resonances are thought to decline with increasing semi-major axis, and the objects themselves may actually spend more time out of resonance than in \citep{yu18}.  Nevertheless, observed distant TNOs including 2008 ST$_{291}$ \citep[conspicuously positioned at the 6:1 resonance:][]{sheppard16_lowe} seem to suggest that the more remote n:1 resonances are indeed populated \citep{pike15,volk16,bannister16b}.

KBOs in the scattered disk with perihelia $\lesssim$ 40 au undergo repeated interactions with Neptune over the life of the solar system that tend to deplete the population via scattering events and ejections \citep{duncan97,nesvorny01}.  However, a small constituency of four objects with $q>$ 50 au and $a\gtrsim$ 100 au including Sedna \citep{sedna}, 2012 VP$_{113}$ \citep{vp113}, 2013 SY$_{99}$ \citep{bannister17_sy99} and Lele\=ak\=uhonua \citep{sheppard19} that are unlikely to be produced in conventional evolutionary models \citep{brasser12} seem to contradict this general trend.  As the orbital poles and longitudes of perihelia of these types of bodies appear to cluster about two opposing directions, \citet{vp113} proposed that the objects are gravitationally shepherded by a distant massive planet in the solar system.  Through this process of secular perturbations from the so-called Planet Nine \citep{batygin16,brown16,sheppard16}, extreme TNO (ETNO) orbits are re-oriented to either oppose (the ``anti-aligned'' population) or match (``aligned'' population) the longitudes of perihelia ($\varpi$) and ascending node ($\Omega$) of the unseen planet.  Multiple follow-on investigations have proposed that the presence of an additional $\sim$5-10 $M_{\oplus}$ planet around $\sim$500-700 au on an eccentric, inclined orbit \citep{batygin16,batygin19_rev} might explain several additional solar system mysteries.  These include a number of highly inclined and even retrograde detected \citep{kv42,kt19} TNOs \citep{batygin16b,becker18}, the 6$\degr$ solar obliquity \citep{bailey16,gomes17}, and the aforementioned high perihelia of scattered disk objects \citep{batygin_morby17,li18,khain18}.  Nevertheless, it is still unclear whether the clustering itself is statistically significant \citep{brown16,sheppard16,brown19} given the observational biases \citep{vp113,shankman17,brown17,kavelaars19}, or simply the consequence of small number statistics \citep{sheppard19,kaib19,clement20_p9,nappier21}.

Similar to the dynamical interplay of resonant TNOs with Neptune, ETNOs are susceptible to MMR capture by the hypothetical Planet Nine \citep{malhotra16}.  In particular, \citet{becker17} found that clones of known ETNOs can abruptly transfer between MMRs with Planet Nine through rapid shifts in their semi-major axes.  Subsequent investigation in \citet{khain20} found that small perturbations from Neptune (even for perihelion passages $\gtrsim$50 au) are responsible for this ``resonant hopping'' phenomenon.  These results represent somewhat of a paradigm shift in the Planet Nine discussion as they suggest that obtaining a comprehensive picture of ETNO dynamics must involve considering the complete Neptune-Planet Nine system.

In this paper we explore the dynamics of the distant n:1 resonances with Neptune \citep[e.g.:][]{lan19} in the presence of Planet Nine.  Our work is motivated by the recent detection of two objects (2007 TC$_{422}$ and 2015 KE$_{172}$) likely inhabiting Neptune's remote 9:1 resonance at $a \simeq$ 130 au \citep{volk18_9_1}.  While previous studies of the outer Kuiper Belt's resonant structure and stability \citep{kaib16,nesvorny16_50au,yu18} have largely focused on KBOs with $a \lesssim$ 100 au (out to the vicinity of the 6:1 resonance), our work considers the complete range of n:1 resonances out to the 14:1.  Moreover, the discovery of 2015 KE$_{172}$ in the 9:1 resonance with a relatively large perihelia ($q =$ 44.1 au) is mildly inconsistent with simulations of Neptune's smooth migration phase following the giant planet instability \citep{nesvorny12,nesvorny16_50au} that suggest such high-$q$ objects would be scattered and lost.  Thus, it is still unclear whether objects in Neptune's furtherest n:1 resonances are primordial or coincidental transient captures.  Our work seeks to understand whether the evolution of particles in the more remote n:1 MMRs changes when Planet Nine is included in the problem.   

\section{Methods}

\subsection{Numerical Simulations}
\label{sect:meth_sim}
Following previous studies \citep{kaib16,nesvorny16_50au,khain20,clement20_p9}, our dynamical investigation involves a suite of numerical simulations utilizing the $Mercury6$ Hybrid integration package \citep{chambers99}.  The dispersal of the primordial Kuiper Belt during the giant planet instability \citep{levison08,nesvorny12} is believed to be primarily responsible for the capture of objects in the distant n:1 resonances with Neptune.  The infant Kuiper Belt itself (also referred to as the remnant ``planetesimal disk'') likely possessed two distinct components: a massive inner region between Neptune and $\sim$30 au \citep[necessary to stop Neptune's outward migration at the correct semi-major axis:][]{gomes04}, and a lower-mass tail extending out to $\sim$45 au.  The massive component likely sourced the various resonant TNO populations and the scattered disk, while the lower-mass region is required to explain the cold classical population of bodies with semi-major axes between $\sim$40-50 au observed today \citep[for a more in depth review see:][]{morby20}.  

After nebular gas dispersal, Neptune migrated smoothly for some time \citep{nesvorny12,bat12,nesvorny15a,nesvorny15b} before experiencing a step-change in semi-major axis during the giant planet instability \citep{deienno17}, followed by a terminal epoch of smooth migration towards its modern orbit at $a_{N}=$ 30.1 au \citep{nesvorny16_grainy,kaib16}.  The precise radial range, migration smoothness and time extent of these various evolutionary phases are still debated.  Indeed, various factors such as Neptune's eccentricity evolution, the duration and depth of planetary encounters during the instability, and the cold classical belt's formation location also play important roles in shaping the modern Kuiper Belt \citep{gomes18,volk19,nesvorny20_i_kb,nesvorny21_e_nep}.  Of additional relevance to our present manuscript, encounters with $\sim$Pluto-mass bodies lead to an evolution of Neptune's semi-major axis that is grainy rather than perfectly smooth.  Such a migration pattern is successful in numerical simulations \citep{nesvorny16_grainy,kaib16} at replicating the modern constituencies of resonant and non-resonant objects.  Indeed, this terminal migration phase is primarily responsible for the capture of objects in the distant n:1 resonances after their semi-major axes and eccentricities are elevated during the giant planet instability \citep{levison08,nesvorny12}.  

As we are particularly interested in maximizing statistics by studying large swarms of remote, resonant TNOs, we begin our numerical investigation at this ultimate epoch of Neptune's radial migration.  In the same manner, we intentionally deviate from the current consensus finding that the temporal evolution of $a_{N}$ was grainy rather than smooth.  In order to generate a large sample of potentially resonant TNOs, we initialize our simulations with Neptune at it's presumed post-jump semi-major axis $a_{n}=$ 28.5 au \citep{deienno17}.  We then artificially force Neptune's residual migration with a fictitious drag force \citep[e.g.:][]{lee02} and 30 Myr e-folding timescale \citep{kaib16}.  Once Neptune reaches its modern semi-major axis, we terminate the fictitious forces, and continue the simulation up to $t=$ 1 Gyr.  All of our integrations include the four giant planets on their modern orbits, utilize a 50.0 day time-step, and remove objects that attain heliocentric distances in excess of 1000 au.  In half of our simulations we include Planet Nine on the preferred orbit proposed by \citet[][$m=$ 5.0 $M_{\oplus}$, $a=$ 500 au, $e=$ 0.25, $i=$ 20.0$\degr$]{batygin19_rev}.  Thus, we perform two separate integrations of each unique set of initial conditions: one with Planet Nine and one without.  In all simulations considering an external perturber, the additional planet is initialized on its presumed modern orbit and is not subjected to artificial migration forces at any point in the integration.  Each individual simulation is designed to study a particular n:1 resonance.  In contrast to studies designed to understand the Kuiper Belt's complex assembly \citep[where fewer particles are scattered into the more distant resonances; e.g.:][]{nesvorny16_50au}, this allows us to easily compare the different resonances with an equal number of initial bodies.  Therefore, we perform 40 simulations (20 which include Planet Nine and 20 that do not) for each individual n:1 resonance between the 3:1 and 14:1 (480 total integrations).  In each simulation, we initialize 500 TNOs with semi-major axes uniformly distributed between the initial value of $a_{res}$ (i.e. the value for $a_{N}=$ 28.5) and the present day location, and perihelia randomly dispersed between 30-40 au\footnote{Note that this selection of initial semi-major axes induces an artificial asymmetry in the final distribution of our TNOs about each n:1 resonance center.  On the sunward side of each resonance, Neptune's migration strands a constituency of TNOs that escaped resonance capture \citep[e.g.:][]{kaib16}.  Conversely, the higher-$a$ extent within the specific width of each resonance \citep[e.g.:][]{lan19} is likely over-filled when compared to the lower-$a$ region as a result placing no objects inside of $a_{res}$ initially.}  Inclinations are selected in a manner designed to replicate the inferred inclination structure of the modern Kuiper Belt \citep[see also:][]{brown01,clement20_p9}:
 \begin{equation}
 	f(i) = sin(i)exp\Big(-\frac{i}{2\sigma_{i}^{2}}\Big)
 	\label{eqn:fi}
 \end{equation}
For our present manuscript, in an effort to maximize the probability of resonant capture, we set $\sigma_{i}$ to 1.0$\degr$ \citep{brown01,kaib16}.  In future work, we plan to consider higher-inclination KBOs.  The remaining angular orbital elements are drawn randomly from uniform distributions of angles

It is worth emphasizing that our simulations are rather fictitious in the sense that they are not meant to study the dispersal of objects in the primordial Kuiper Belt towards the distant n:1 resonances.  Instead, we tune our computations in order to create a large population of potentially resonant objects in the respective n:1 MMRs.  For this reason, we neglect Neptune's primordial migration and instability-induced jump \citep{nesvorny12,deienno17}, model the final approach to $a_{N}$ with unrealistically smooth semi-major axis evolution \citep{nesvorny16_grainy,kaib16} and disregard the effects of stellar encounters \citep[e.g.:][]{fernandez00,rickman08} and the galactic tide \citep[which are not expected to significantly perturb our TNOs:][]{heisler86,dones04,kaib09}.  Thus, our simulations should be interpreted purely as a stability analysis of the various distant resonances in the presence of Planet Nine, and not as a robust evolutionary study of the region.  However, it is still an open-ended question whether objects in the remote n:1 MMRs with Neptune such as 2007 TC$_{434}$ and 2015 KE$_{172}$ are primordial, or intermittent captures \citep{volk18_9_1}.  As our simulations account for Neptune's residual migration phase that would presumably have been responsible for primordial capture of resonators, we do comment on the feasibility of this mechanism in section \ref{sect:kb_i}.

\subsection{Resonance Identification}
\label{sect:res}

\begin{figure}
	\centering
	\includegraphics[width=.5\textwidth]{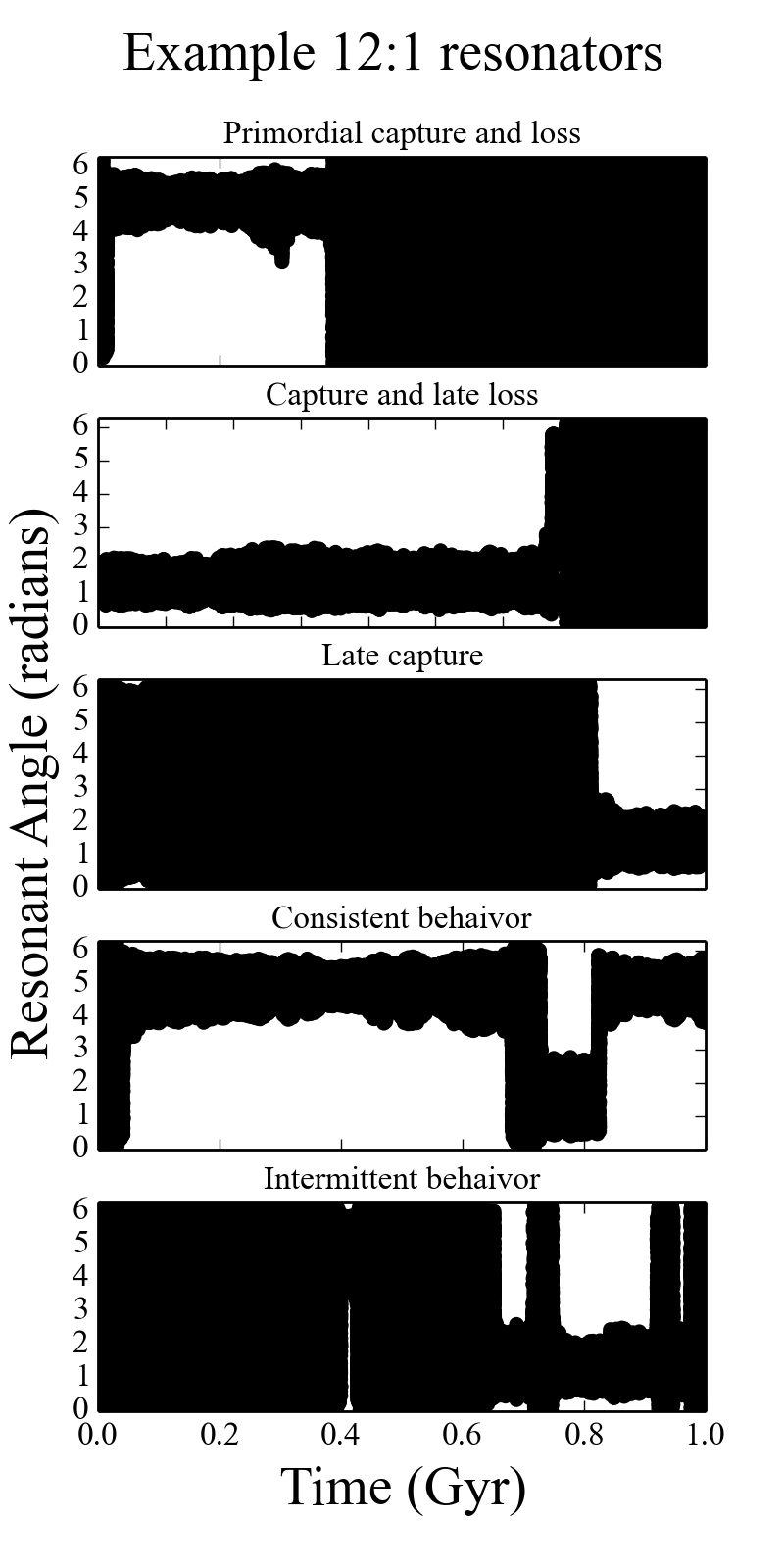}
	\caption{Example evolution of the resonant angles (equation \ref{eqn:res}) of five objects in our simulations studying Neptune's 12:1 MMR characterized as resonant by our classification pipeline (section \ref{sect:res}).}
	\label{fig:resang}
\end{figure}

We leverage a simple statistical pipeline to quickly identify objects librating about resonant angles of the form:
\begin{equation}
	\phi = p \lambda_{TNO} - q \lambda_{N} - r \varpi_{TNO} - s \varpi_{N}
	\label{eqn:res}
\end{equation}
\begin{equation}
	p=q+r+s
\end{equation}
where $\varpi$ is the longitude of perihelion and $\lambda$ is the mean longitude of each object.  For each particle in our simulations, we compute the various possible resonant angles over 1 Gyr from simulation outputs every 10,000 years, and calculate the circular standard deviation \citep{mardia72} for each time window of interest.  We confirmed via visual inspection of sample distributions that objects with a resonant angle standard deviation $\lesssim$2.0 exhibit significant libration for at least 20$\%$ of the given epoch \citep[see also:][]{clement17}.  Figure \ref{fig:resang} depicts five different example evolutionary sequences that are characterized by our classification system as resonant.  While our statistical pipeline clearly overlooks resonant behavior that is excessively intermittent or brief, we elect this approach as it allows us to consistently and unambiguously compare and contrast the dynamical evolution within each respective n:1 resonance.

\section{Results and Discussion}

\subsection{Depletion of the distant resonances by Planet Nine}
\label{sect:sect1}

\begin{figure}
	\centering
	\includegraphics[width=.5\textwidth]{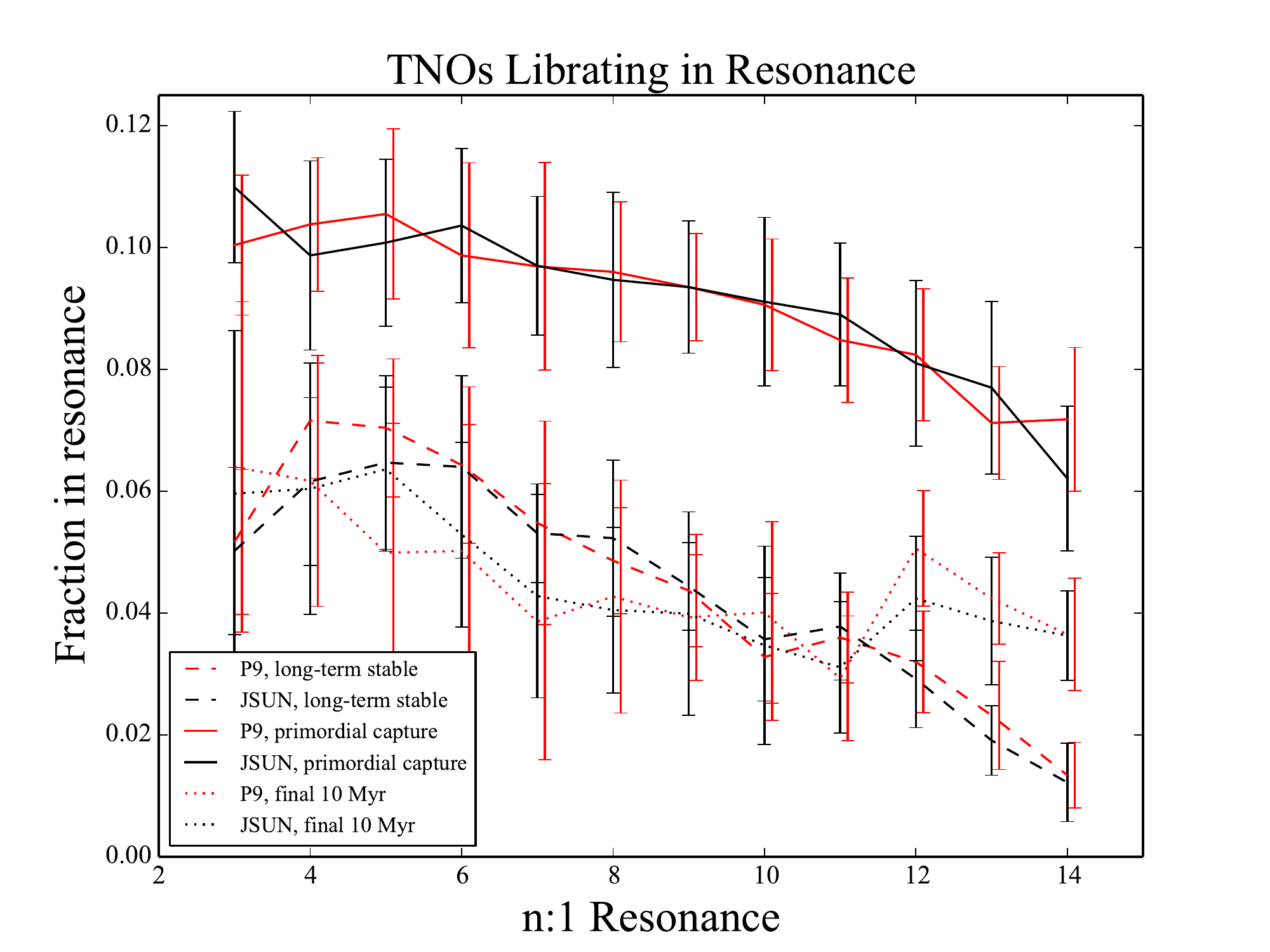}
	\caption{Relative population of TNOs librating (section \ref{sect:res}) in the various n:1 resonances investigated by our simulations.  The solid lines represent primordial captures during Neptune's forced migration phase.  Conversely, the dashed lines represent objects that exhibit significant resonant behavior over the 1 Gyr duration of the simulation.  Finally, the dotted lines denote objects that are resonant within the final 10 Myr of the simulation.  Simulations that include Planet Nine are plotted in red, and the black lines represent simulations that did not.  Thus, the acronyms \textit{JSUN} and \textit{P9} (used throughout our manuscript) indicate simulations that include only the four known giant planets and those that include Planet Nine, respectively.  The error bars are proportional to the 1-$\sigma$ deviation between various individual simulations.}
	\label{fig:n_1_ai}
\end{figure}

To first order, our simulations indicate that Planet Nine's presence does not affect the \textit{fraction of objects trapped in} Neptune's remote n:1 resonances at any point in the simulation.  However, the distant perturber \textit{significantly erodes the population} of TNOs in the vicinity of the furthest n:1 resonances (e.g.: the 12:1, 13:1 and 14:1; characterized in greater detail in section \ref{sect:ero}).  In particular, our simulations investigating the 12:1 resonance lose 20$\%$ of TNOs via ejections or perihelia detachment from Neptune (q $>$ 50 au) when Planet Nine is included in the calculation, compared to just 11$\%$ when it is not.  This trend continues for the 13:1 (22$\%$ loss in 1 Gyr with Planet Nine and 12$\%$ without) and the 14:1 (24$\%$ versus 13$\%$) resonances.  Planet Nine's tendency to deplete distant KBOs in this manner is well documented.  Indeed, studies of the external perturber's influence on objects in the remote Kuiper Belt consistently find that distributions of test particles initialized with uniform distributions of semi-major axes and perihelia are depleted in number by around an order of magnitude over the life of the solar system \citep{khain18,saillenfest20_review,clement20_p9}.  However, our simulations evince a rich and complex dynamical environment within Neptune's most remote MMRs when Planet Nine is incorporated in the calculation.  In particular, we find that the hypothetical planet's tendency to raise the perihelia of distant TNOs and detach them from the realm of Neptune's influence operates efficiently both within, and outside of the furthest n:1 MMRs.  As resonant libration in these distant MMRs is rather infrequent and intermittent regardless of the inclusion of Planet Nine, the fact that our 1 Gyr Planet Nine simulations studying the 13:1 and 14:1 resonance with Neptune lose nearly half of the TNOs initially in the region suggests that such objects would be quite scarce if the solar system contained an additional massive planet beyond the Kuiper Belt.

Figure \ref{fig:n_1_ai} plots the fraction of objects flagged as resonant by our identification algorithm (section \ref{sect:res}) for three different time windows of interest.  The solid lines reflect objects captured in resonance during Neptune's simulated migration phase, which we refer to as ``primordial captures.''  We discuss the evolution of these objects in greater detail in section \ref{sect:kb_i}.  The dashed lines in Figure \ref{fig:n_1_ai} compile objects that we determine to be stable in resonance for $\sim$100 Myr timescales.  We make this determination by computing the circular standard deviation of the cumulative range of simulation outputs.  Thus, these data effectively represent objects that exhibit significant and consistent resonant behavior for at least 200 Myr of our 1 Gyr simulations.  Example resonant angle evolutionary schemes for these types of objects are plotted in Figure \ref{fig:resang}.  With this in mind, we acknowledge that the dashed lines in Figure \ref{fig:n_1_ai} combine multiple varieties of TNO evolution within these MMRs.  In particular it is difficult to disentangle objects that intermittently bounce in and out of resonance \citep[e.g.:][]{lykawka07} from those that do not. Indeed, our resonance identification method is biased in favor of objects that are consistently stable in resonance (e.g.: panels 2 and 4 of Figure \ref{fig:resang}) by virtue of consolidating a large range of time outputs, and as a consequence of the inherent limitation of our prescribed simulation output cadence of 10,000 years.  As the dashed lines in Figure \ref{fig:n_1_ai} evidence a strong trend of long-term stable objects becoming scarcer within Neptune's n:1 resonances with increasing $n$, it is reasonable to question whether this trend in fact demonstrates increasingly intermittent and sporadic resonant libration at higher $n$.  For this reason, we repeat our analysis on the final 10 Myr time window of our simulations (dotted lines).   The flattened slope of the dotted lines at high $n$ thus suggests that the strong trend of smaller resonant populations with increasing $n$ is the result of increasingly intermittent resonant libration.  Thus, our simulations lead us to conclude that objects in Neptune's most remote n:1 resonances (regardless of the inclusion of Planet Nine) tend to bounce in and out of resonance and different resonant arguments, and exhibit significantly less consistent resonant behavior than objects in the nearest MMRs.

\begin{figure*}
	\centering
	\includegraphics[width=.9\textwidth]{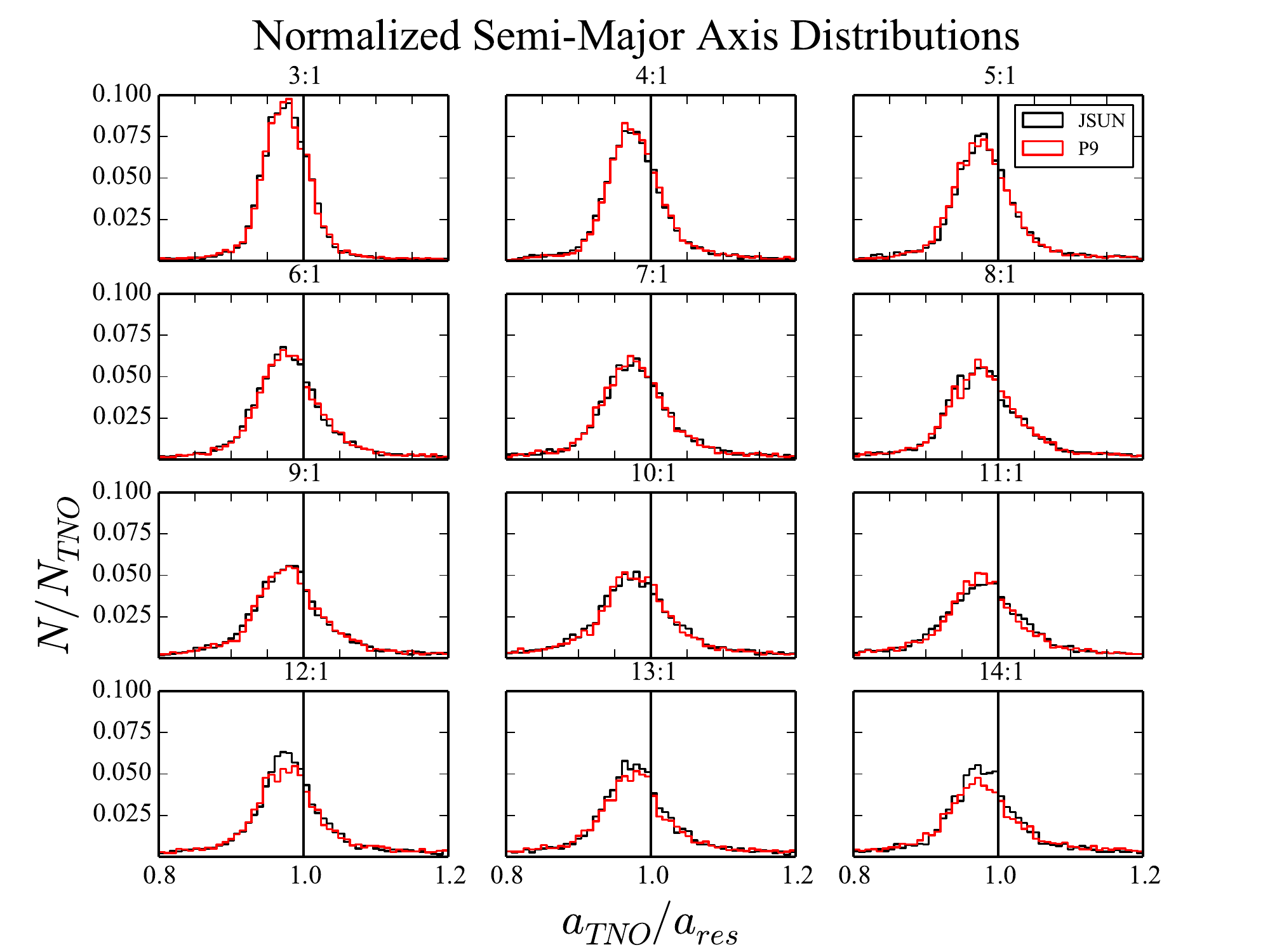}
	\caption{Final normalized semi-major axis distributions about the respective n:1 resonances with Neptune in our various simulation sets.  Each distribution is normalized to one by the final number of surviving objects in the simulation.}
	\label{fig:tails}
\end{figure*}

In general, Figure \ref{fig:n_1_ai} does not reveal meaningful differences in TNO resonant behavior between our simulation sets that include a Planet Nine model, and those that do not.  However, it is important to note that the dashed and solid curves in figure \ref{fig:n_1_ai} plot the total fraction of objects \textit{remaining in the simulation} that are resonant with Neptune.  As the simulations incorporating a Planet Nine model finish with significantly fewer particles than those without an external perturber, the total number of resonant TNOs in our Planet Nine simulations is necessarily smaller.  This implies that Planet Nine's secular perturbations efficiently erode \textit{both} the resonant and non-resonant populations.  We explore the mechanisms governing this depletion of resonant TNOs further in section \ref{sect:ero}.

Figure \ref{fig:tails} plots the distribution of objects about the exact resonant center in our various simulation sets (note that each individual distribution is normalized to sum to one).  One noticeable feature in this figure that is common among all the various resonances investigated in our study is that the peak of the distribution of object semi-major axes is slightly interior to that of the precise resonant center as computed via Kepler's laws.  There are two reasons for this discrepancy.  The first is that Neptune's migration phase fossilizes a ``tail'' of objects that avoid resonant capture \citep[for a more detailed discussion see:][]{kaib16}.  Thus, this artificial peak essentially corresponds to the center of the initial distribution of objects in our simulations.  The second, albeit less profound reason is that the allowable range of libration zones for highly eccentric orbits shifts slightly, and noticeably widens around the arithmetic resonant center.  \citet{lan19} mapped the stable resonant zones for Neptune's exterior n:1 resonances out to the 10:1, and we direct the reader to their paper for a more detailed discussion of the stable resonant parameter space (in particular their figures 1 and 6).

The second noticeable trend in Figure \ref{fig:tails} is that the center of the high-$n$ distributions is less distinct in our simulations that include Planet Nine (note that each distribution is normalized to one, we discuss depletion in section \ref{sect:ero}).  In particular, our 12:1 Planet Nine simulations possess 12.5$\%$ fewer objects with semi-major axes within 5$\%$ of the resonant center than the counterpart runs that do not include an external perturber.  This trend continues in our 13:1 simulations (14.4$\%$ depression with Planet Nine) and 14:1 integrations (15.9$\%$).  This evidences a correlation that we expand upon in the subsequent section: more efficient erosion of the most distant n:1 resonances in simulations that incorporate Planet Nine. Thus, coupling these results with the trends noted in Figure \ref{fig:n_1_ai}, we conclude that the dominant mechanism operating within our simulations studying the higher-$n$ resonances is Planet Nine efficiently removing objects with systematically shorter resonant ``sticking'' timescales \citep{lykawka07}.  In the same manner, strong Kozai cycles operating near these distant MMRs cycle the eccentricities of these objects rather dramatically, thus making it easier for their perihelia to become detached via secular perturbations from Planet Nine.  We highlight some of the more interesting dynamical evolutionary paths embarked on by these eroded TNOs in appendix \ref{sect:troj}.

\subsection{Erosion Timescale}
\label{sect:ero}

\begin{figure*}
	\centering
	\includegraphics[width=.9\textwidth]{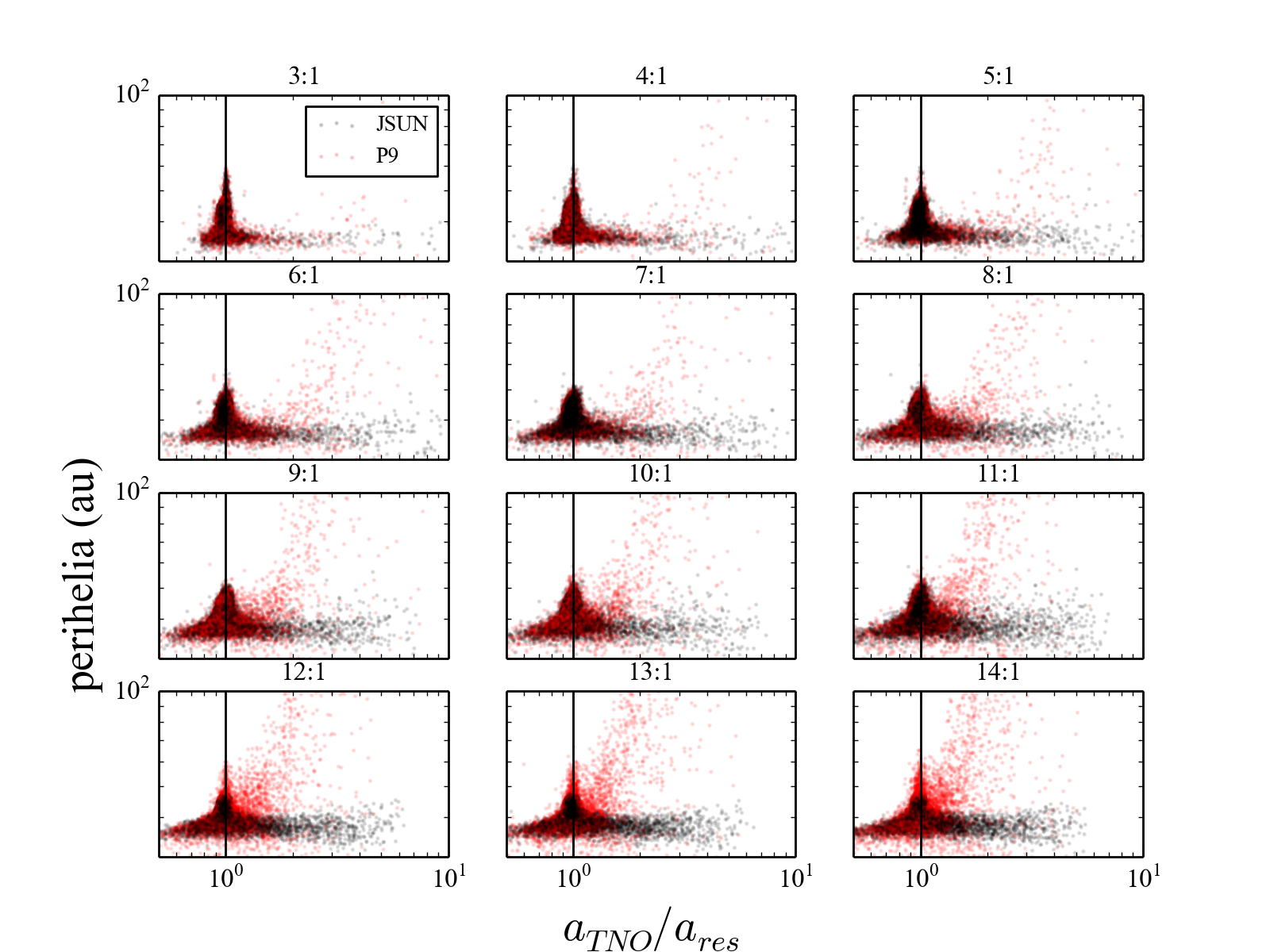}
	\caption{Final perihelia distribution around the respective n:1 resonances with Neptune in our various simulation sets.  Simulations that included Planet Nine are plotted in red and those that did not are plotted in grey.}
	\label{fig:qa_jsun}
\end{figure*}

Our simulations indicate that perturbations from the hypothetical Planet Nine efficiently erode the inventories of TNOs scattered into Neptune's distant n:1 resonances.  Figure \ref{fig:qa_jsun} plots the final $a/q$ distributions in our various simulations sets with and without a Planet Nine model.  In all cases, scattering events with Neptune near perihelia over time tend to disperse each resonances' eccentricity distribution; typically placing objects on orbits with larger semi-major axes.  This process is most active for the more remote, and systematically weaker n:1 resonances \citep{lykawka07}.  A noticeable feature in both figures is the pronounced peak of relatively stable, high-perihelia ($q\gtrsim$ 40 au) orbits near the heart of each resonance.  This perihelia lifting occurs when particles trapped in a MMR with Neptune undergo Kozai oscillations in $e/i$ \citep{gomes05_kozai}.  Otherwise, surviving TNOs remain largely bounded within a range of perihelia ($\sim$30-40 au) where their dynamics remain rather strongly coupled to Neptune.

\begin{figure*}
	\centering
	\includegraphics[width=.49\textwidth]{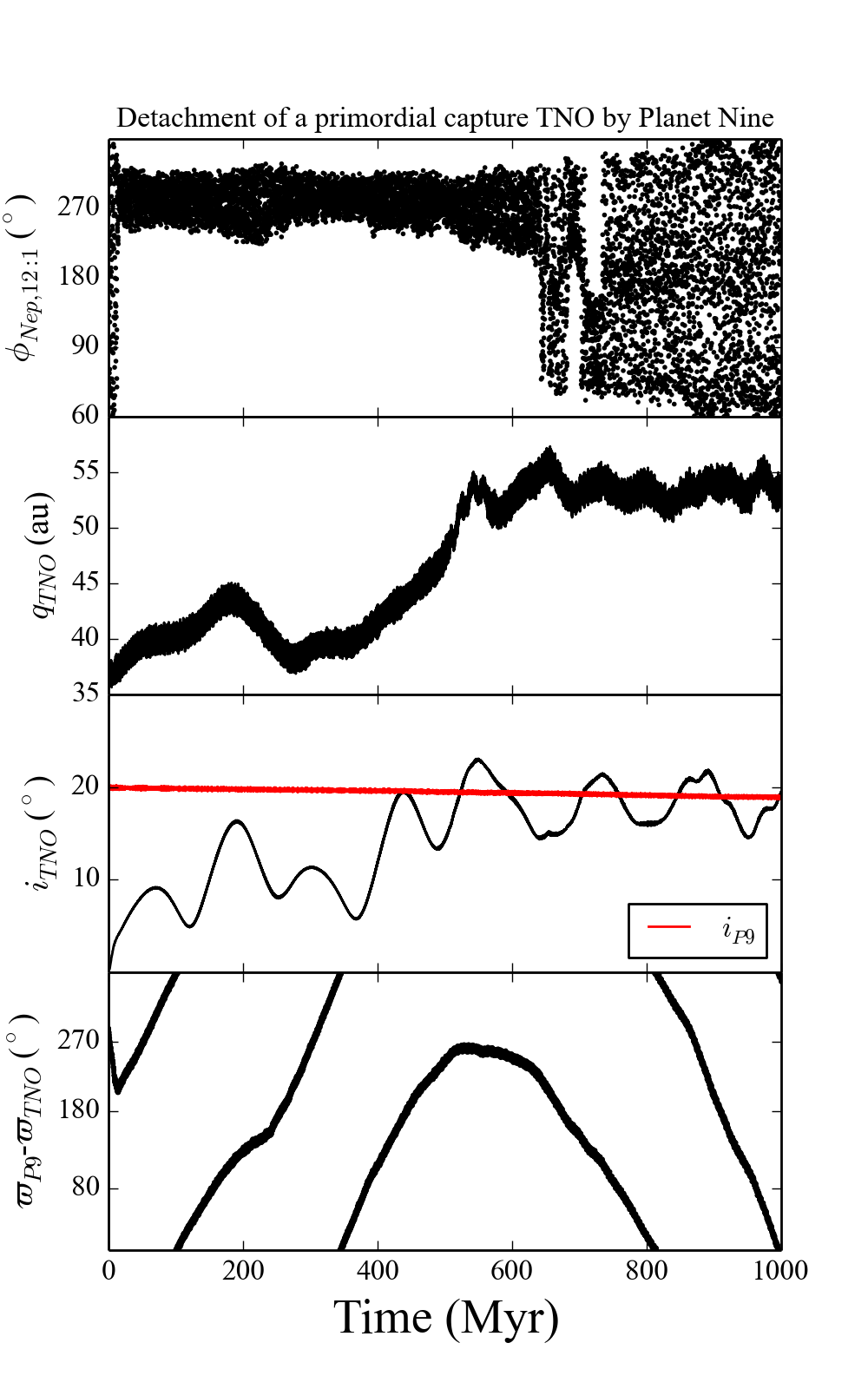}
	\includegraphics[width=.49\textwidth]{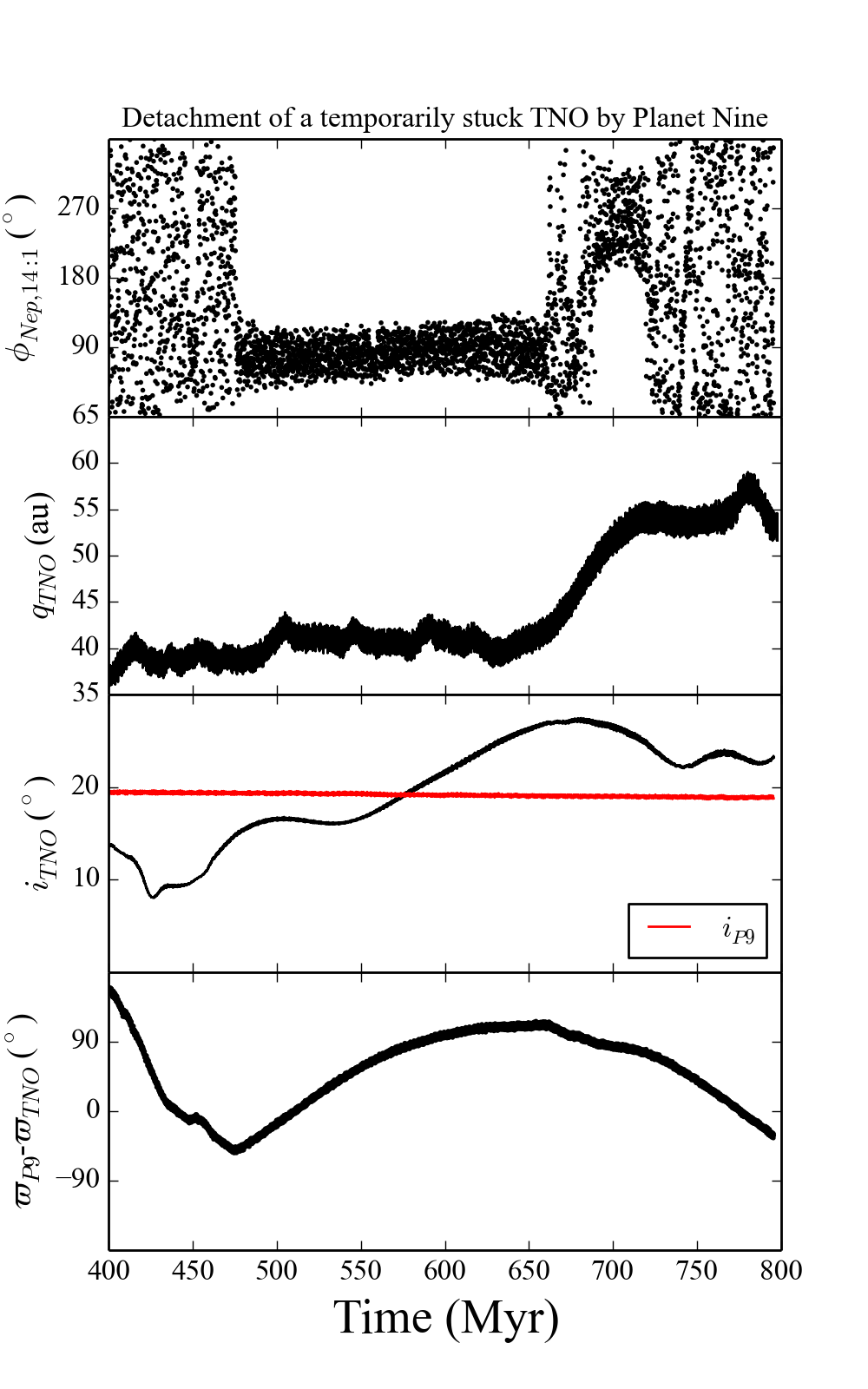}	
	\caption{Two representative example evolutions of resonant TNOs experiencing perihelia detachments as the result of secular perturbations from Planet Nine that ultimately dislodge them from MMRs with Neptune.  The left panel depicts the evolution of a TNO captured in Neptune's 12:1 resonance during the planets' initial migration phase, while the right panel plots a TNO that spontaneously sticks in the 14:1 resonance after $\sim$500 Myr of non-resonant activity.  The top panels plot the respective 12:1 and 14:1 MMR angles with Neptune.  Panels 2-4 depict the evolution of each TNO's perihelia, inclination, and longitude of perihelia with respect to that of Planet Nine, respectively.}
	\label{fig:detach_ex}
\end{figure*}

The picture in Figure \ref{fig:qa_jsun} changes rather substantially when Planet Nine is included in the simulation.  In all cases, an apparent tail of high-$a$ objects with lifted perihelia \citep{sheppard16_lowe,batygin16} is apparent.  In the most distant n:1 resonances, it is clear that this depletion mechanism is operating within the resonance itself.  Thus, objects in our 12:1, 13:1 and 14:1 simulations need not be scattered to larger semi-major axes in order to be efficiently detached from Neptune by Planet Nine.  Figure \ref{fig:detach_ex} plots two example evolutionary histories of TNOs that experience perihelia detachments \textit{while in resonance with Neptune}.  The object depicted in the left panel is captured in the 12:1 MMR during Neptune's migration phase at the beginning of the simulation.  After $\sim$200 Myr of regular resonant evolution characterized by Kozai oscillations in $e$ and $i$, the secular angle describing apsidal confinement with Planet Nine ($\varpi_{P9}-\varpi_{TNO}$) shifts from circulation to libration around 180$\degr$ (an anti-aligned configuration).  Through this process of secular $e-\varpi$ coupling the TNO's perihelia begins to inflate; thus decoupling it from Neptune's influence and removing it from the 12:1 resonance around $t=$ 700 Myr (note that the characteristic period of the secular angle in panel 4 matches that of the object's perihelia evolution in panel 2).  The right panel plots a similar interaction where a TNO sporadically sticks \citep{lykawka07} in Neptune's 14:1 MMR at $t=$ 500 Myr in an aligned configuration with Planet Nine.  Interestingly, in this example it is clear that the object's excited inclination plays an important role in its' detachment around $t=$ 700 Myr (note that the dominant oscillation in $i$ depicted in panel 3 matches that of the secular angle in the bottom panel).  Indeed, Planet Nine's tendency to detach TNOs in this manner is well documented \citep{batygin16b,batygin19_rev}.  However, our simulations suggest that the mechanism operates efficiently within Neptune's distant n:1 MMRs.  In this manner, we expect Planet Nine to significantly erode the inventories of both resonant and non-resonant TNOs with $a\gtrsim$150 au (around the 12:1 MMR.

\begin{figure*}
	\centering
	\includegraphics[width=\textwidth]{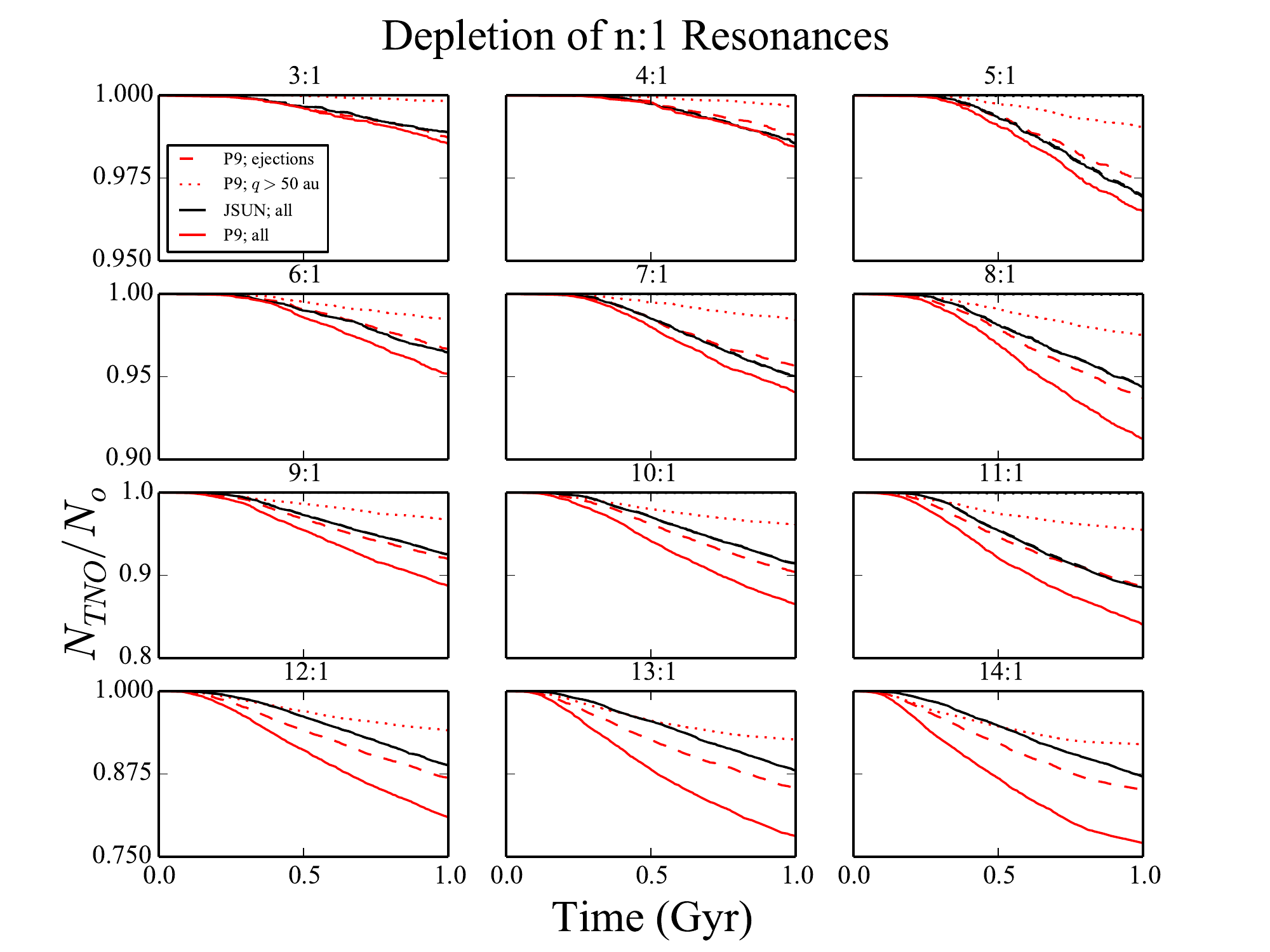}
	\caption{Depletion and detachment of TNOs from Neptune's distant n:1 resonances over our 1 Gyr simulations.  Each sub-panel plots the cumulative results of 40 different simulations studying each particular resonance.  Black lines represent data from simulations that did not include a distant massive planet, and red lines denote the simulations that included Planet Nine.  The dashed lines plot particle ejections ($Q>$ 2,000 au) and the dotted lines track particle detachment from Neptune (the first time an object that is not ejected achieves $q>$ 50 au; note that JSUN simulations essentially produce no detachments).  The solid lines compile both ejections and detachments.}
	\label{fig:deplete}
\end{figure*}

We study the depletion histories of our various resonant populations in Figure \ref{fig:deplete}.  The most simplistic types of particle loss in our simulations to track are ejections.  However, we note that objects considered ejected in our simulations are not always the ones that attain hyperbolic orbits, but also those that reach heliocentric distances in excess of 2,000 au.  This places the most distant $e\approx$1 object tracked by our simulations at 1,000 au; around the exterior 3:1 MMR with Planet Nine.  As this is well within the regime of the inner Oort Cloud, such a delimiting distance will suffice for the purposes of this paper.  While the dashed lines in Figure \ref{fig:deplete} correspond to particle ejections, we are also interested in objects that become detached from Neptune and enter the realm of the ETNOs (q$\gtrsim$ 50 au).  However, Kozai cycles within the MMRs evaluated in our paper can cause objects' perihelia to temporarily exceed 50 au without truly detaching them from Neptune's influence.  The strength and timescale of these variations depend on the initial perihelia, inclination and semi-major axes of the TNOs.  Objects in our simulations with initial perihelia near 40 au and low inclinations ($i<$5$\degr$) typically spend $\gtrsim$15$\%$ of our simulations with $q>$45 au via $\sim$100 Myr timescale oscillations in $e$ and $i$. To isolate TNOs that are significantly decoupled from Neptune's dynamics, we plot the last time objects with a final time averaged perihelia in excess of 50 au (over a 10 Myr interval) have $q<$50 au (dotted lines).  Finally, the solid lines in Figure \ref{fig:deplete} tabulate the cumulative erosion rates from both ejections and detachments.
Figure \ref{fig:deplete} clearly demonstrates substantial erosion of Neptune's distant TNO populations when Planet Nine is included in the calculation.  While perihelia detachments seldom occur in integrations without Planet Nine (for instance we note only five such events in our entire batch of 10:1 simulations without Planet Nine), they continue steadily throughout the simulation in those incorporating an additional planet.  In this manner, detachments are largely responsible for the deviation in net depletion after 1 Gyr in the two sets of simulations (and the different distributions in Figure \ref{fig:tails}).

\subsection{Comparison with the observed population}

\begin{figure*}
	\centering
	\includegraphics[width=.8\textwidth]{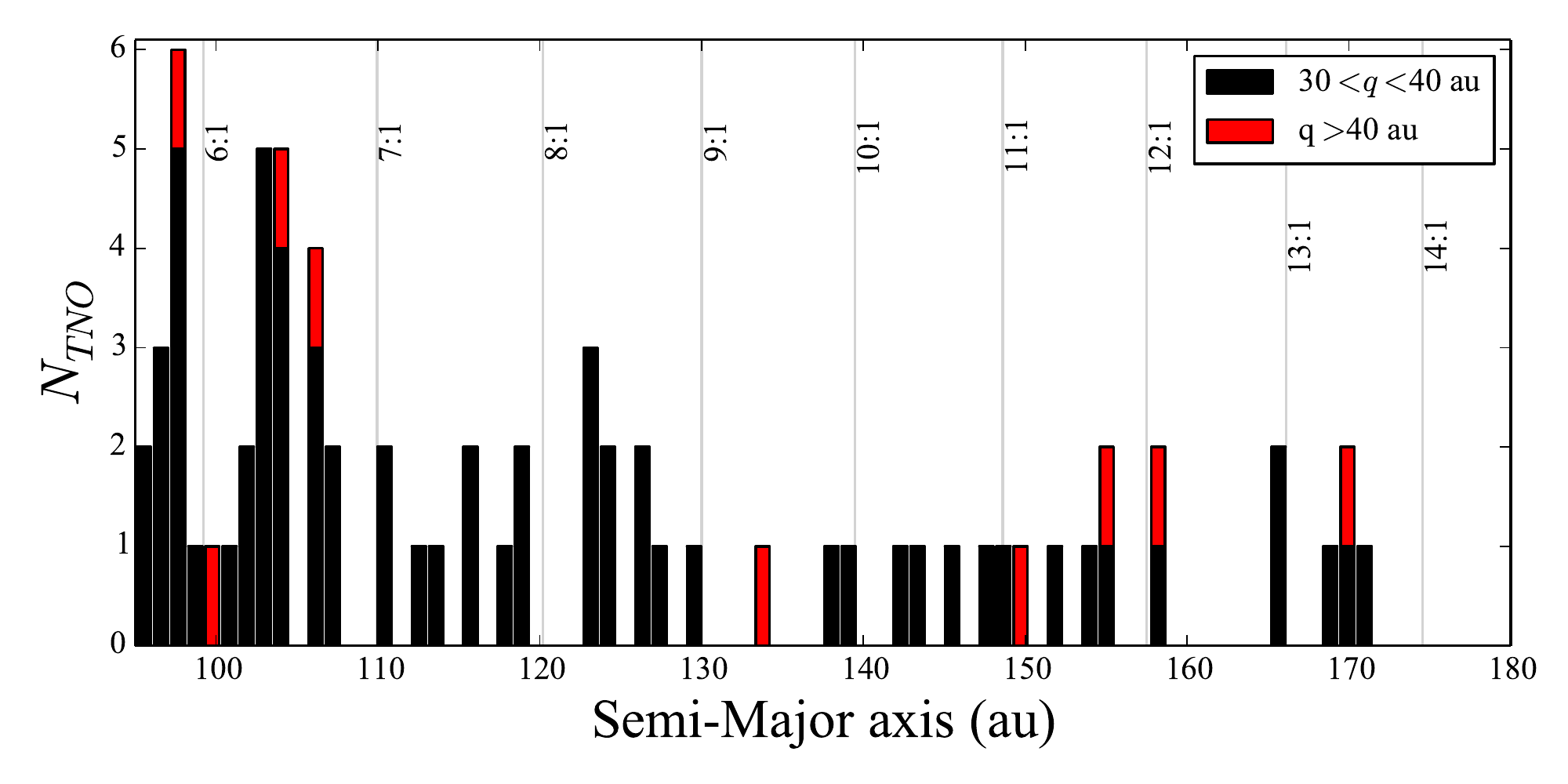}
	\caption{Semi-Major Axis distribution of all TNOs in the vicinity of Neptune's distant n:1 resonances according to the $MPC$ database queried on 28 December 2020.}
	\label{fig:semi}
\end{figure*}

The perihelia-lifting and corresponding particle erosion mechanism demonstrated in our simulations that include Planet Nine is well explored in the literature \citep{batygin16,batygin_morby17,khain18,batygin19_rev}.  Thus, the more interesting result of our study is in the potential of observed objects (Figure \ref{fig:semi}) in the vicinity of Neptune's remote resonances to either confirm or deny Planet Nine's existence.  We acknowledge that it would not be particularly worthwhile to compare the ratios of observed populations near the successive n:1 MMRs with the results of our study in order to test the validity of the Planet Nine hypothesis.  Indeed, our simulations do not account for the transfer of objects between semi-major axis bins, nor do they predict how populated each resonance should be initially as we do not completely model the dispersal of the primordial Kuiper Belt \citep{kaib16,nesvorny16_grainy}.  However, it is clear that Planet Nine efficiently removes resonant objects from the most remote n:1 resonances (neglecting repopulation via objects with initially lower semi-major axes).  While TNOs beyond the 9:1 MMR should be rare regardless of the presence of additional massive bodies in the outer solar system, objects near the 12:1, 13:1 and 14:1 should be exceedingly rare if the Planet Nine hypothesis is correct.  Currently, there are nine known objects (see Figure \ref{fig:semi}) with 30 $<q<$ 40 au, and semi-major axes between 150 and 180 au (roughly in the vicinity of the 12:1, 13:1, 14:1 resonances).  Conversely, their are 11 objects with semi-major axes near the 9:1, 10:1 and 11:1 (125-150 au).  Thus, to first order there does not appear to be a significant dip in the population of KBOs out as far as the 14:1.  

Table \ref{table:reals} summarizes the known objects with 100 $<a<$ 200 au and $q>$ 30 au that are potentially resonant with Neptune.  To quickly develop these determinations, we generate 100 clones of each object by imposing small deviations of $\delta a<$ 1.0 au and $\delta e<$ 0.01 on each TNO's orbit as reported in the \textit{MPC} database).  We then integrate each clone in the presence of the modern solar system for 4 Gyr.  Next, we analyze the time series of possible resonant arguments (equation \ref{eqn:res}) for the nearest n:1 resonance (based on the initial semi-major axis) using the same circular standard deviation methodology described in section \ref{sect:res} and a rolling time window of 10 Myr (each window of analysis overlaps the previous by 5 Myr).  Through this process, we determine the fraction of clones that are currently in resonance, the median time they remain stuck in resonance (from the present), the average fraction of the total integration they librate within the appropriate MMR, and the median time in which the objects permanently depart the n:1 resonance.  It is worth recognizing here that the vast majority of our clones exhibit resonant behavior of some kind during the simulation \citep[as is the expected behavior in the region:][]{gallardo06,lykawka07}.  Many transiently stick in an n:1 resonance at some point in the future, a sizable fraction are currently in a higher-$m$ resonance (i.e. n:2, n:3, and so on), and others experience sticking-epochs in multiple resonances.  Thus, we only list TNOs in table \ref{table:reals} which were resonant in at least 20$\%$ of our clone simulations.  Notably, we identify three objects (2004 PB$_{112}$, 2015 DW$_{224}$, 2013 TV$_{158}$) potentially locked in Neptune's 7:1 resonance.  While a full dynamical characterization of the resonant behavior of these extreme objects is beyond the scope of this manuscript, we present this small batch of simulations to provide the reader with a first order characterization of the population of observed distant resonators.  Furthermore, we note that only 9$\%$ of 2007 TG$_{434}$ clones \citep[an object argued as likely inhabiting the 9:1 resonance in][]{volk18_9_1} librate in the 9:1 in our simulations.  Conversely, our results for 2015 KE$_{172}$ (also determined to be in the 9:1 MMR in that work) are broadly consistent with that of \citet{volk18_9_1}.  Finally, it is worth remarking that our characterization of 2014 JW$_{80}$ and 2014 OS$_{394}$ as likely in the 10:1 and 11:1 resonances would make them the most distant known resonant TNOs.  As the uncertainty in each of these objects' orbits is extremely small (\textit{MPC} Uncertainty Factors of 0 and 1, respectively), we plan to fully characterize the likelihood of their orbits being securely resonant with an appropriate number of simulations (i.e.: greater than 100 clones) in future work.

The same general trends discussed in sections \ref{sect:sect1} and \ref{sect:ero} hold for our simulations of observed TNO clones.  While the fraction of objects in resonance and the respective sticking times do not change appreciably when Planet Nine is included in the computation (given the small number statistics involved with analyzing just 100 clones), the median last time interval of resonant libration is systematically later in simulations without an additional planet.  This again supports our conclusion that Planet Nine efficiently raises the perihelia of both the resonant and non-resonant TNOs near Neptune's remote n:1 MMRs.  Indeed, only 21 of our 2014 JW$_{80}$ clones and 32 of our 2014 OS$_{394}$ clones near the 10:1 and 11:1 resonances are lost via ejections or perihelia detachments in simulations without Planet Nine.  Conversely, the number of particle losses increase to 32 and 50 (for 2014 JW$_{80}$ and 2014 OS$_{394}$, respectively) when the external perturber is included.  While it would be naive to argue that this speaks against the Planet Nine hypothesis in any meaningful way, particularly given the aforementioned caveats, the fact that modern surveys are beginning to probe and characterize a region of trans-Neptunian space that should be rather strongly influenced by Planet Nine implies that these objects might one day provide an intriguing constraint on the planet's existence and properties.

\begin{table*}
	\centering
	\begin{tabular}{c c c c c c}
	\hline
	Object & Resonance & $N_{res}/N_{clone}$ & $t_{stick}$ (Myr) & $t_{stick}/t$ & $t_{loss}$ (Myr) \\
	\hline
	2013 VS$_{46}$ & 6:1 & 29 (30) & 465 (695) & 0.43 (0.54) & 3940 (3900) \\
	2004 PB$_{112}$ & 7:1 & 19 (19) & 10 (10) & 0.048 (0.053) & 3820 (3880) \\
	2015 DW$_{224}$ & 7:1 & 43 (30) & 55 (32.5) & 0.34 (0.28) & 3350 (2775) \\
	2013 TV$_{158}$ & 7:1 & 18 (21) & 48 (40) & 0.30 (0.35) & 2095 (1885) \\
	2015 RQ$_{281}$ & 8:1 & 34 (33) & 38 (75) & 0.168 (0.12) & 3610 (2975) \\
	2015 KE$_{172}$ & 9:1 & 41 (37) & 100 (145) & 0.28 (0.34) & 3110 (2710) \\
	2018 VO$_{35}$ & 9:1 & 25 (27) & 10 (5) & 0.029 (0.033) & 905 (570) \\
	2014 JW$_{80}$ & 10:1 & 51 (45) & 15 (20) & 0.079 (0.090) & 2600 (2370) \\
	2014 OS$_{394}$ & 11:1 & 33 (31) & 30 (15) & 0.058 (0.077) & 3425 (2820) \\
	\hline
	\end{tabular}
	\caption{Summary of resonant sticking times for objects with significant sticking timescales in 4 Gyr integrations of all known TNOs with 100 $<a<$ 200 au and $q>$ 30 au.  Results from integrations that included Planet Nine are given in parenthesis.  The columns are as follows: (1) The object's MPC designation, (2) the n:1 resonance the object likely currently inhabits, (3) the fraction of all clones in an n:1 resonance at the beginning of the simulation, (4) the median current sticking timescale (i.e.: the median time before resonant libration ceases), (5) the fraction of the total integration the object is stuck in resonance, and (6) the median time of permanent loss from resonance (i.e.: the last time output where the object is librating in resonance).}
	\label{table:reals}
\end{table*}

\subsection{Resonance Hopping}
\label{sect:hop}

\begin{figure}
	\centering
	\includegraphics[width=.5\textwidth]{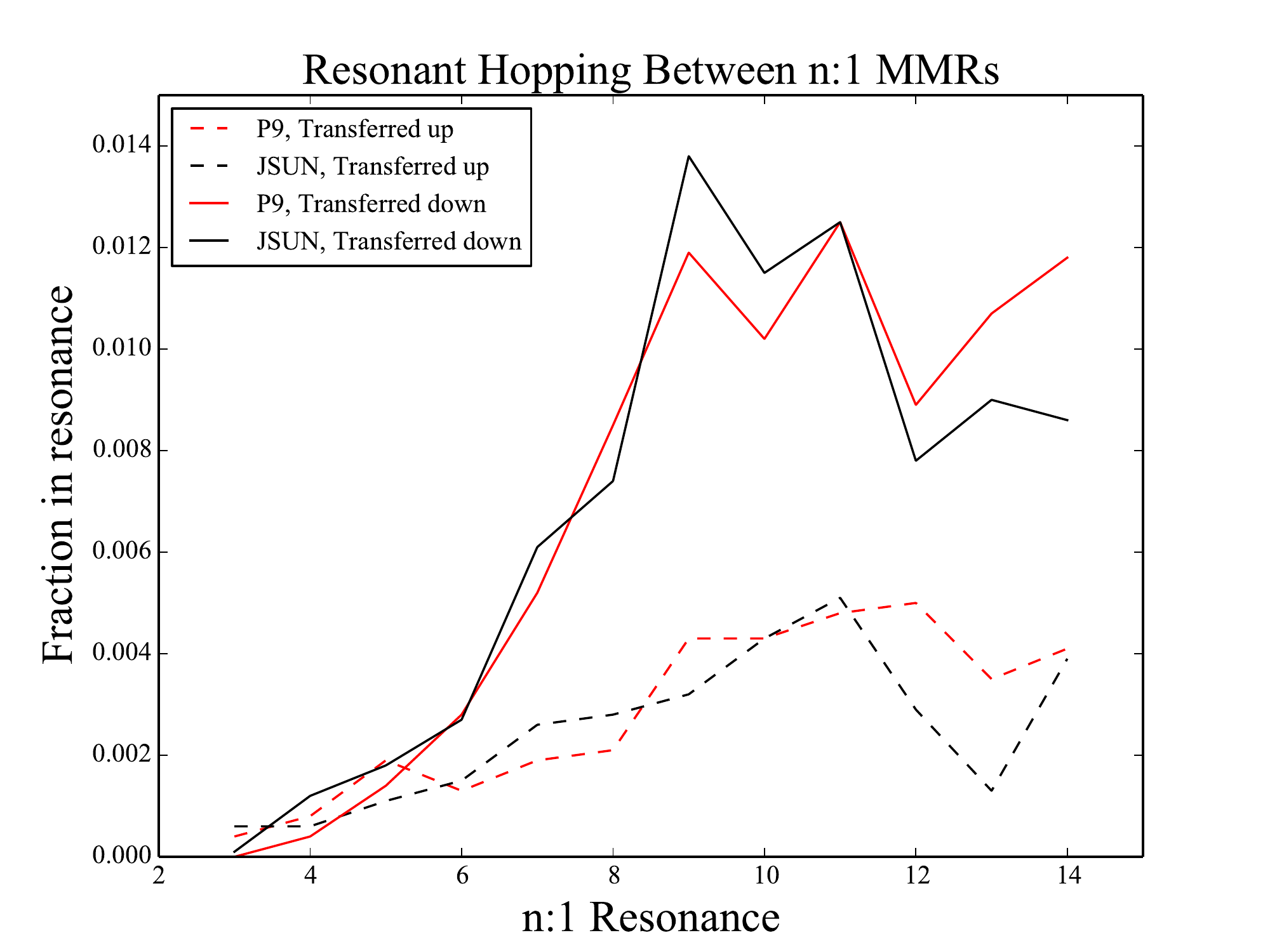}
	\caption{Fraction of objects in our various simulation sets transferred between successive n:1 MMRs.  The solid lines plot objects that begin in a lower n:1 MMR and are transferred to a higher ((n+1):1) resonance during the integration.  Conversely, the dashed lines represent particle exchange from n:1 MMRs to (n-1):1.  simulations that include Planet Nine are plotted in red, and the black lines represent simulations that did not.  As the variance in results between runs within a given simulation sets is quite broad, for ease of interpretation we do not include 1-$\sigma$ deviations as in Figure \ref{fig:n_1_ai}.}
	\label{fig:up_down}
\end{figure}

In the final sections of our manuscript (\ref{sect:hop}, \ref{sect:kb_i} and appendix \ref{sect:troj}), we return the discussion to our original simulations investigating TNO evolution in successive n:1 resonances described in \ref{sect:meth_sim}.  We note numerous instances of transfer between successive n:1 MMRs with Neptune in these simulations.  Figure \ref{fig:up_down} plots the total fraction of objects exchanged between the n:1 resonances investigated in our various simulation sets with the respective (n-1):1 (dashed lines) and (n+1):1 (solid lines) MMRs.  Clearly, transfer to a higher-n resonance is far more common than downward exchange.  This is not surprising given the tendency of scattering interactions with Neptune to lift the semi-major axes of our distant TNOs \citep{duncan97,nesvorny01}.  Interestingly, beyond the 8:1 resonance, upward transfer is quite common, and the total number of objects that spend time in a higher-n resonance is relatively comparable to the number that librate in the resonance of interest (see Figure \ref{fig:n_1_ai}) in each simulation set.  This again speaks to the increasingly intermittent and sporadic mode of resonant behavior we pointed out in section \ref{sect:sect1} that is typical at high-n.  We also note a minor trend of increased transfer between resonances when Planet Nine is included in the integration.  While this phenomenon is quite rare within our sample of simulations, it appears to be the result of Planet Nine's tendency to efficiently lift TNO's perihelia and semi-major axes via secular perturbations.

\subsection{Primordial capture}
\label{sect:kb_i}

We also scrutinized the evolution of TNOs captured in n:1 resonances during Neptune's migration phase.  It is clear from Figure \ref{fig:n_1_ai} that the capture efficiency slightly decreases with increasing $n$.  Coupled with the fact that  the population of the more distant semi-major axis bins during the epoch of planetesimal disk dispersal is expected to be less efficient than for the closer n:1 MMRs \citep{yu18}, one might expect primordial captures to be quite rare in the more distant resonances.  However, when we compare the objects in our simulations librating in resonance in the first 100 Myr (within Neptune's simulated migration phase) with those resonant in the final 10 Myr, we find a roughly equal number of stable primordial captures throughout our various simulation sets (albeit with a substantial degree of scatter).  Regardless of the n:1 resonance investigated in our simulations, or the Planet Nine model invoked, individual simulations tend to finish with $\sim$5 primordial captures still locked in resonance.  However, we were unable to decipher any meaningful correlations between the orbits of the primordial captures, and those entrapped later in the simulation.  Thus, we conclude that it is possible for primordial n:1 resonant captures to still exist in the distant regions of trans-Neptunian space today regardless of Planet Nine's existence.  However, these objects are quite rare, and likely do not dominate the population of any given n:1 resonance.

\subsection{Resonant Objects with Planet Nine possessing $a\lesssim$ 100 au}
\label{sect:new_p9_res}

\begin{figure}
	\centering
	\includegraphics[width=.5\textwidth]{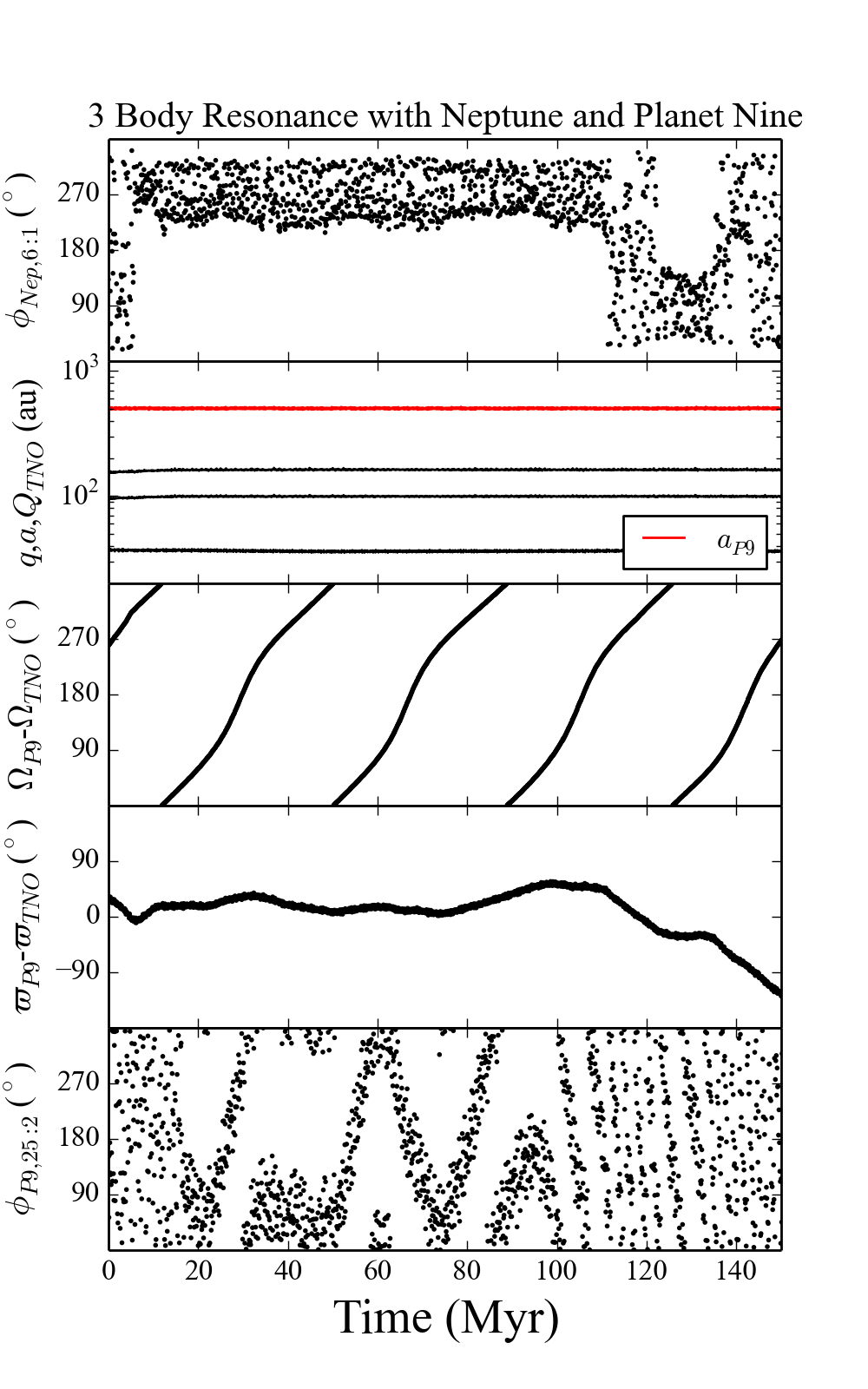}
	\caption{Example 3 body resonant behavior between a TNO ($q=$ 37 au; $a=$ 93 au), Neptune and Planet Nine.  The top panel depicts the dominant resonant angle (equation \ref{eqn:res}) for the Neptune-TNO 6:1 MMR.  The second panel plots the static behavior of the TNO's perihelion, semi-major axis and aphelion.  The third and fourth panels depict the evolution of the particle's longitudes of ascending node and perihelion with respect to that of Planet Nine.  The bottom panel show the relevant resonant angle for the 25:2 TNO-Planet Nine resonance.}
	\label{fig:3body}
\end{figure}

We note three instances of objects with $a<$ 100 au that inhabit MMRs with Planet Nine (specifically in the 15:1, 25:2 and 29:3; see Figure \ref{fig:p9_res}).  All three possess low inclinations ($i<$ 2.5$\degr$) and perihelia between 40 and 45 au.  Thus, they exist in a curious regime where their semi-major axes are rather low and their perihelia are only moderately detached compared to those of objects typically considered in the Planet Nine discussion \citep[e.g.:][]{batygin16,brown19,sheppard19,clement20_p9}, however their dynamics still couple weakly to the distant planet.  Thus, this rare class of objects in our simulations represent a cross-over regime between the Neptune-TNO and Planet Nine-TNO problems.  Interestingly, one of these TNOs exhibits a brief $\sim$ 100 Myr period of resonant and near-resonant libration with both Neptune's 6:1 resonance and Planet Nine's 25:2 resonance while at a perihelia of 39.0 au.  We plot the evolution of this three body resonance in Figure \ref{fig:3body}.  While it is not clear if the 25:2 resonant angle is truly librating, during the applicable time of resonant interaction the TNO remains aligned in perihelia with Planet Nine.  Thus, there is a strong secular component to resonant coupling with Planet Nine.  Unfortunately, this region of overlapping influence between Neptune and Planet Nine is not well constrained observationally in the solar system.  According to the $MPC$ database, there are currently just three cataloged objects with similar orbits (90 $<a<$ 100 au and 40 $<q<$ 45 au): 2008 ST$_{291}$, 2010 ER$_{65}$ and 2015 GB$_{56}$, all of which have rather large inclinations of 20-30$\degr$ that can be explained by the Kozai resonance and MMR interactions with Neptune \citep{sheppard16_lowe}.  Thus finding a low inclination object with a semi-major axis near 100 au and perihelion near 40 au may be a sign of an object in resonance with Planet Nine.

\section{Conclusions}

In this paper we presented numerical simulations investigating the stability and dynamical evolution of trans-Neptunian objects (TNOs) in Neptune's successive n:1 resonances (specifically those between the 3:1 and 14:1; semi-major axes of $\sim$60-180 au).  By considering two separate models, one that included a distant massive planet \citep[the so-called hypothetical Planet Nine:][]{vp113,batygin16} and one that did not, we concluded that objects with semi-major axes that place them near Neptune's 12:1, 13:1 and 14:1 resonances are significantly eroded by perturbations from Planet Nine (note that here we define erosion as both particle ejections and perihelia detachments; $q\gtrsim$50 au).  

Specifically, our simulations investigating the 13:1 MMR lose 22$\%$ of TNOs when Planet Nine is included in the problem, compared to 12$\%$ when it is not.  Similarly, our 14:1 runs incorporating a Planet Nine model erode 24$\%$ of TNOs in 1 Gyr, as opposed to just 13$\%$ in runs without a distant perturber.  Of perhaps greatest interest, we find that secular perturbations from the massive perturber efficiently lift the perihelia of resonant, and non-resonant TNOs near these MMRs.  Thus, the detection of a significant number of distant objects near these MMRs with low perihelia might one day provide a useful constraint on the Planet Nine hypothesis.  Indeed, modern surveys are beginning to characterize this region of orbital parameter space \citep[e.g.:][]{volk18_9_1,sheppard19}.  

Through an additional batch of integrations fashioned to investigate the evolution of all known TNOs with $a>$ 100 au, we identify 9 objects potentially locked in a distant n:1 MMR with Neptune.  By analyzing the evolutions of these clones, we diagnose the orbits of 2014 JW$_{80}$ and 2014 OS$_{394}$ as potentially belonging to the 10:1 and 11:1 MMR populations, respectively.  While the scope of our analysis was insufficient to fully characterize these objects' dynamics, our determination makes them the most distant known objects that are potentially resonant with Neptune.  Additionally, our study identifies four objects likely inhabiting the 7:1 (2004 PB$_{112}$, 2015 DW$_{224}$ and 2013 TV$_{158}$) and 8:1 (2015 RQ$_{281}$) MMRs.  Given the difficulties involved with detecting more objects similar to the ETNOs and inner Oort Cloud objects used to infer Planet Nine's existence \citep{vp113,sheppard16,bannister27,bannister18,sheppard19}, further study and observation of the region thought to be dominated by n:1 resonant TNOs with Neptune \citep{lykawka07,nesvorny16_50au} would be quite useful.

Our simulations illuminated a rich dynamical environment in the trans-Neptunian region when Planet Nine is included in the calculation.  Indeed, the dynamical behavior of particles evolving between the Sun-Neptune-TNO problem and the Sun-Planet Nine-TNO regime in our integrations is quite vibrant and complex.  Of greatest interest, our simulations produce three examples of mean motion resonant capture by Planet Nine at exceedingly low semi-major axes (a$\lesssim$100 au).  In all three cases, the objects posses low inclinations and moderate perihelia.

While our results are intriguing in terms of demonstrating potentially meaningful ways in which the Planet Nine hypothesis might be constrained in the future, we remind the reader that our simulations are extremely simplistic and limited in terms of their ability to capture the complex formation and early dynamical evolution of the trans-Neptunian regime.  Specifically, we design our study to maximize our sample of potentially resonant TNOs, and therefore do not accurately model the dispersal of the primordial Kuiper Belt \citep{nesvorny16_grainy,kaib16} that presumably generated these objects.  This allows us to directly compare the dynamics within the successive n:1 resonances, however this analysis is plagued by the obvious degeneracy that arises from the fact that our simulations are not geared to deduce \textit{how populated the remote resonances should be.} Thus, future work should aim to self-consistently assemble the distant ($a\gtrsim$ 50 au) Kuiper Belt in the presence of Planet Nine.  Finally, our work is limited in that we only consider one potential orbit for the hypothetical Planet Nine \citep{batygin19_rev}, and limit the duration of the majority of our numerical integrations to 1 Gyr.  Thus, our present manuscript serves as a first order proof-of-concept of how the Planet Nine hypothesis might be indirectly constrained via objects in the vicinity of Neptune's distant n:1 MMRs.  As the population of such detected objects continues to increase, future investigations should expand on the ideas presented here with an eye towards more thoroughly confining the feasibility of the Planet Nine hypothesis.

\section*{Acknowledgments}

We thank John Chambers, Nate Kaib and an anonymous reviewer for useful comments and insight that greatly improved the manuscript and the presentation of the results.  The work described in this paper was supported by Carnegie Science's Scientific Computing Committee for High-Performance Computing (hpc.carnegiescience.edu).

\appendix
\section{MMR capture by Planet Nine}
\label{sect:troj}

In this appendix we expound upon the discussion of TNO resonance capture by Planet Nine from section \ref{sect:new_p9_res}.  It is important to note that our numerical experiments are not designed to accurately estimate the hypothetical resonant population of Planet Nine as we do not self-consistently model the dispersal of the primeval Kuiper Belt, and the corresponding formation of the scattered disk.  Nonetheless, our simulations produce a plethora of interesting dynamical evolutions as TNOs traverse between the well-studied region of the giant planets' influence, towards more exotic high-$e$, high-$i$ resonant interactions with Planet Nine.  The following sections briefly detail certain noteworthy evolutions and classes of TNO orbits that might provide meaningful constraints on Planet Nine's existence as more objects with detached perihelia are discovered in the future.

The potential of scattered ETNOs to become captured in both interior and exterior resonances with Planet Nine has been noted before \citep{malhotra16,batygin_morby17,becker17,khain18}.  We searched for objects exhibiting significant libration in resonance with Planet Nine over the 1 Gyr duration of our simulations using the same methodology described in section \ref{sect:res}.  To search for resonances, we consider all possible n:m resonant angles within 10$\%$ of the object of interest's period ratio with Planet Nine with $m\leq$ 8 (equation \ref{eqn:res}).  While it is known that ETNOs ``hop'' between different higher order resonances, the sticking timescales observed in numerical integrations are rather lengthy given the large timescales for secular eccentric oscillations in the region \citep{khain20}.  Thus, we assess our resonant detection algorithm to be an adequate means of separating objects caught in resonance from those that are not.  In total, we detect 61 total resonances between our TNOs and Planet Nine.  Figure \ref{fig:p9_res} summarizes the different resonances detected, as well as the simulation set in which the resonant TNO originated.  It is clear that the majority of the resonant captures are objects initially in the vicinity of Neptune's more remote n:1 resonances, particularly those beyond the 10:1.  However, we find at least one instance of resonance capture in each set of simulations.  Consistent with previous work, the stronger MMRs of the form n:1 and n:2 are fairly prevalent in our simulations \citep{becker17,bailey18}.  However, we also find the first and second order resonances (e.g.: 7:6, 6:5, 4:3, 7:5) in the vicinity of $a_{P9}$ to be well represented as well.  While our simulations do not track high eccentricity objects beyond the 3:1 (by virtue of their extreme aphelia flagging them as ejections), to first order, the interior resonances do tend to be significantly more populated than the exterior resonances.  

\subsection{Trojan captures}

\begin{figure*}
	\centering
	\includegraphics[width=.96\textwidth]{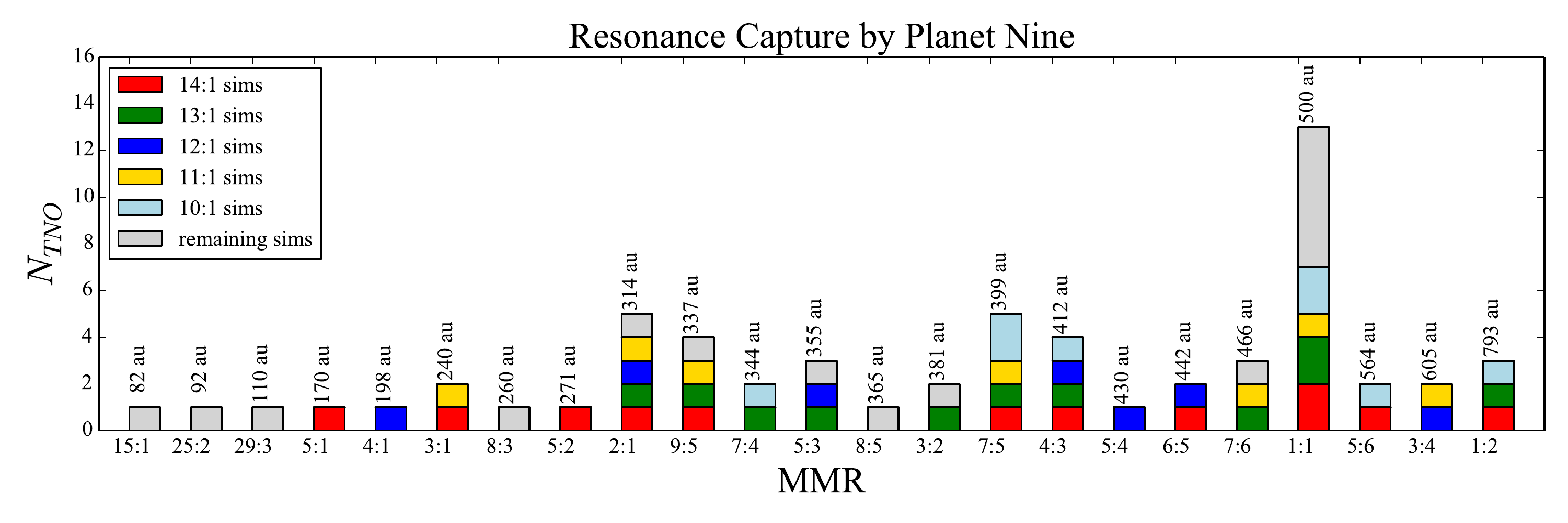}
	\caption{Summary of TNOs captured in MMRs with Planet Nine in our various simulation sets.  The color of each bar corresponds to the n:1 resonance with Neptune that the simulation investigated (section \ref{sect:meth_sim}).  The semi-major axis of the center of each resonance is provided above each bar.}
	\label{fig:p9_res}
\end{figure*}

By far, the most populated resonance with Planet Nine in our simulations is the 1:1 \citep[a similar result was found in][]{batygin_morby17}.  Specifically, we detect 13 instances where one of our TNOs is captured in a co-orbital configuration with Planet Nine.  As our simulations considering a distant massive planet collectively investigate 120,000 individual particles, we acknowledge that this phenomenon is quite rare.  However, only a small number of our TNOs become detached from Neptune and thus enter the regime where their dynamics are governed by Planet Nine (section \ref{sect:ero}).  While only a fraction of those objects with elevated perihelia are captured in MMRs with Planet Nine \citep[e.g.:][]{becker17,khain20}, it is apparent from Figure \ref{fig:p9_res} that capture in the 1:1 MMR is relatively common.  Thus, we conclude our manuscript by briefly commenting on the interesting dynamical behavior exhibited by these trojan TNOs.

\begin{figure}
	\centering
	\includegraphics[width=.5\textwidth]{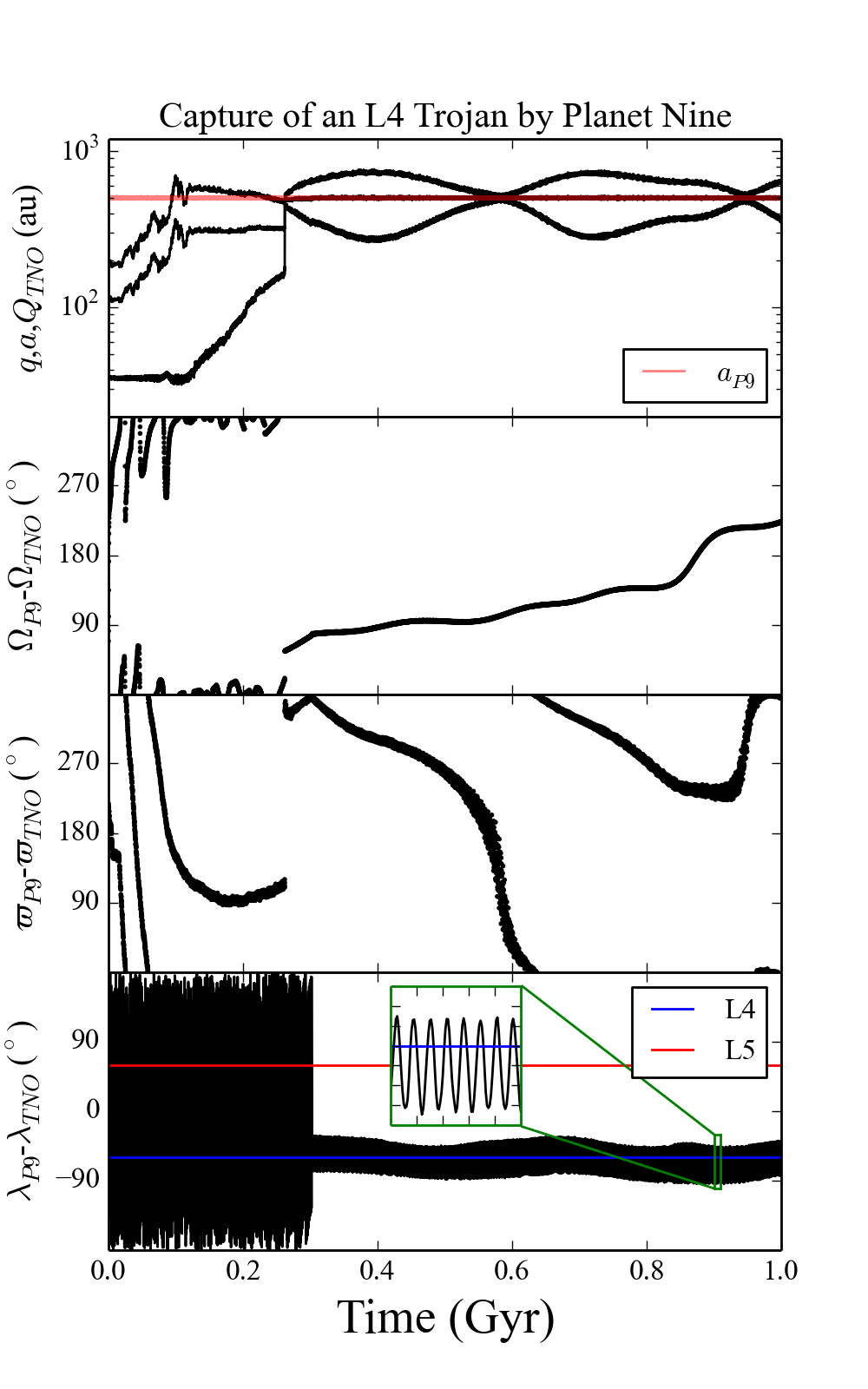}
	\caption{Evolution of a trojan TNO captured by Planet Nine in a simulation studying Neptune's 8:1 resonance.  The top panel plots the TNO's perihelion, semi-major axis and aphelion in black, and Planet Nine's semi-major axis in red.  The second panel depicts the alignment of the TNO's longitude of ascending node ($\Omega_{TNO}$) with that of the distant massive planet over the duration of the simulation.  The third panel displays the evolution of the TNO's longitude of perihelion ($\varpi_{TNO}$) with respect to Planet Nine's.  The transition from circulation to opposition of this critical argument of the $\nu_{9}$ resonance around $t=$100 Myr is responsible for raising the TNO's perihelia and detaching it from Neptune's influence.  The bottom panel depicts the difference between the mean longitudes of the Planet Nine and the TNO.  The inserted panel (green) zooms in on a 10 Myr section of the simulation.  The blue and red lines indicate the locations of L$_{4}$ and L$_{5}$, respectively.}
	\label{fig:troj_l4}
\end{figure}

Of our 13 1:1 resonant TNOs, only five are stable from the point of capture to the end of the integration.  The remaining eight are lost at some point in the simulation with a median time in resonance of 245 Myr.  Intriguingly, only three of the 13 instances of 1:1 resonant behavior occur about Planet Nine's L$_{4}$ and L$_{5}$ Lagrange points.  An example of such a capture is plotted in Figure \ref{fig:troj_l4}.  The second panel depicts how nodal perturbations from Planet Nine contribute to incrementally exciting its eccentricity.  Through this process, the TNO attains an orbital orientation that aligns with that of Planet Nine in $\Omega$, and opposes it in $\varpi$.  As the TNO's semi-major axis begins to inflate, it reaches a configuration where it begins to periodically encounter Planet Nine while at aphelion starting around $t=$ 121 Myr.  Over the next $\sim$ 100 Myr, the object undergoes a series of 211 encounters with the distant massive planet ($r_{enc} \sim$15-80 au).  Finally, a close approach of $r=$ 0.029 au at $t=$ 261 Myr significantly perturbs the TNO's orbit; thereby leading to its capture at L$_{4}$.
\begin{figure}
	\centering
	\includegraphics[width=.5\textwidth]{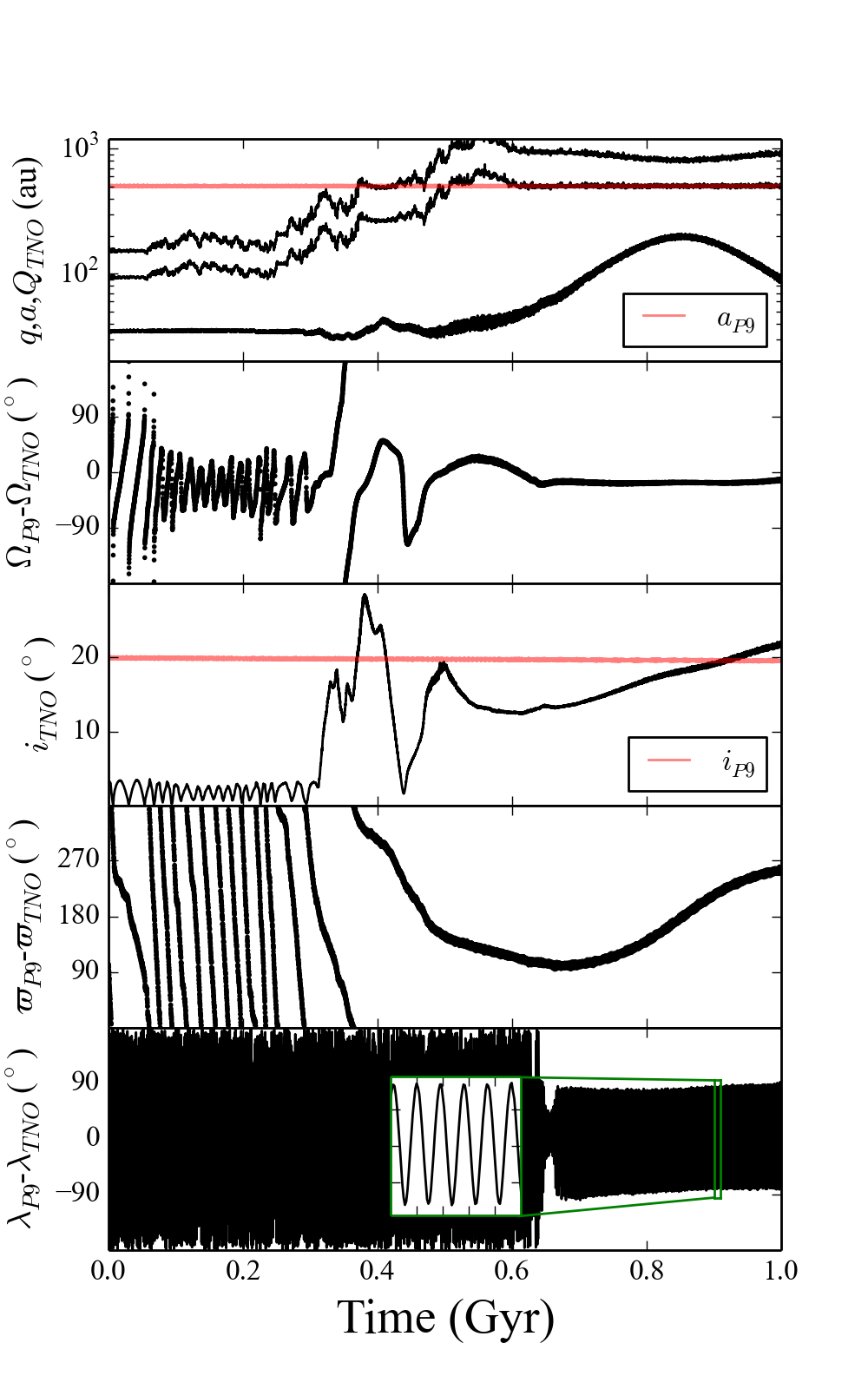}
	\caption{Same as Figure \ref{fig:troj_l4}, except here for a secularly shepherded trojan captured by Planet Nine in a simulation studying Neptune's 6:1 resonance.  The top panel plots the TNO's perihelion, semi-major axis and aphelion in black, and Planet Nine's semi-major axis in red.  The second panel depicts the alignment of the TNO's longitude of ascending node ($\Omega_{TNO}$) with that of the distant massive planet over the duration of the simulation.  The third panel plots the inclinations of the TNO (black) and Planet 9 (red) over the duration of the simulation.  The fourth panel displays the evolution of the TNO's longitude of perihelion ($\varpi_{TNO}$) with respect to Planet Nine's.  The transition from circulation to opposition of this critical argument of the $\nu_{9}$ resonance around $t=$400 Myr is responsible for raising the TNO's perihelia and detaching it from Neptune's influence.  The bottom panel depicts the difference between the mean longitudes of the Planet Nine and the TNO.  The inserted panel (green) zooms in on a 10 Myr section of the simulation.}
	\label{fig:troj}
\end{figure}
The majority of our trojan TNOs exhibit a more exotic phase protection mechanism \citep[e.g.:][]{robutel06} than the lower-eccentricity example depicted in Figure \ref{fig:troj_l4}.   Co-orbital motion in the circular restricted problem is constrained by the perturbing body's Jacobi constant \citep{schubart64}.  In this regime, the familiar ``tadpole'' orbits about L$_{4}$ or L$_{5}$, as well as horseshoe orbits traversing L$_{3}$, L$_{4}$ and L$_{5}$ are bounded by an annulus with interior and exterior radii at points L$_{1}$ and L$_{2}$.  At high eccentricity and inclination, the classic analytical picture changes substantially as the Jacobi constant is no longer sufficient to define allowable trajectories for the small body.  In this high-eccentricity regime, the libration center for phase-protected orbits shifts to the vicinity of the planet.  Often referred to as the so-called ``retrograde satellite'' class of orbits \citep{henon70,apostolos00}, these dynamically excited trojans' mean anomalies concentrate near that of the perturbing planet (e.g.: Figure \ref{fig:troj}), rather than leading or trailing by 60$\degr$ as is the case for tadpole orbits about L$_{4}$ or L$_{5}$.  Thus, in the frame of the planet, the particle appears to exhibit quasi-periodic motion; hence the moniker "retrograde satellite.''  This type of dynamical behavior is well-studied in the solar system, particularly for eccentric Near-Earth Objects \citep[NEOs, e.g.:][]{apostolos00,nesvorny02_co_orb} that exhibit epochs of exotic co-orbital motion in numerical integrations.  Perhaps the most famous of this class of objects is the temporary Earth trojan (3753) Cruithne with $e=$ 0.515 and $i=$ 19.8$\degr$ \citep{namouni99}.

Ten of our trojan-capture TNOs are best classified as retrograde satellites \citep{henon70}.  However, as a result of the lengthy orbital periods, slow secular precession timescales, and extreme eccentricities, the dynamics recorded in our simulations are quite different from those displayed by a Sun-Earth-NEO retrograde satellite.  Figure \ref{fig:troj} plots an example of a trojan capture of this variety by Planet Nine.  Nodal secular perturbations from the remote planet detach the TNO's perihelion from Neptune and elevate its inclination \citep{batygin16,batygin16b} beginning around $t=$ 300 Myr (panels 1, 2 and 3 or Figure \ref{fig:troj}).  Once captured as a co-planar, eccentric co-orbital object with $i_{TNO} \approx i_{P9}$ and $\Omega_{TNO} \approx \Omega_{P9}$ around $t=$ 600 Myr, the particles motion is governed by two dominant cycles.  The first is a slow periodic oscillation about a longitude of perihelion that is opposed to that of Planet Nine (panel 4) with an amplitude of $\sim$180$\degr$ and a characteristic period of $\sim$700 Myr.  Within this cycle, the object's mean anomaly oscillates around that of Planet Nine with a period of $\sim$2 Myr and an amplitude of $\sim$180$\degr$ (panel 5).  Thus, the phase protection mechanism is predominantly secular in nature (note that the least-extreme perihelion and aphelion configuration in panel 1 around $t=$ 850 Myr occur $\varpi_{P9}-\varpi_{TNO} \simeq$ 180$\degr$).  In other words, the ability of our high-eccentricity TNO to avoid close encounters with Planet Nine is intrinsically driven by the e-$\varpi$ coupling depicted in panels 1 and 4.  For further information regarding this coupling within the 1:1 MMR we direct the reader to figure 13 of \citet{batygin19_rev} and the accompanying text.

We extended the integrations of our ten retrograde satellite TNOs to $t=$ 4 Gyr and found that only the one plotted in Figure \ref{fig:troj} still inhabited the same resonant structure at the end of the simulation.  Therefore, as is the case with the exotic classes of NEO trojan objects, the Planet Nine co-orbital captures in our simulations tend to only produce quasi-stable configurations.  Indeed, our secularly shepherded retrograde satellite co-orbitals exhibit large-amplitude oscillations in $e$ that lead them to spend large periods of time at perihelia near 50 au.  During this time of repeated excursions towards the region of Neptune's influence, periodic energy exchanges with Neptune can eventually dislodge the TNO from resonance \citep[e.g.:][]{khain20}.  Nonetheless, these trojans do tend to exhibit exceedingly long dynamical lifetimes compared to similar exotic trojans in the inner solar system by virtue of the lengthy orbital and secular precession periods in the region.  Thus, they manifest as the most common class of resonant capture by Planet Nine in our simulations (Figure \ref{fig:p9_res}).  Moreover, the 1:1 resonance with Planet Nine is not particularly susceptible to the resonance-hopping phenomenon that is thought to transfer objects between various higher-order MMRs \citep{becker17}.  These results suggest that, given a sufficient number of detected ETNOs with aligned longitudes of ascending node and perihelion, a 1:1 resonance-based search for the hypothetical Planet Nine might be possible \citep[e.g.:][]{malhotra16,bailey18}.  At first glance, the conspicuously oriented orbits of Senda ($a=$ 484 au, $i=$11.9$\degr$; $\Omega=$105$\degr$) and 2013 RA$_{109}$ ($a=$ 479 au, $i=$12.4$\degr$; $\Omega=$144$\degr$) seem to be potentially consistent with this picture, and might belong to a hypothetical population of Planet Nine co-orbitals \citep[indeed, the semi-major axes and inclinations of these objects are quite similar to the preferred Planet Nine model of ][also utilized in this manuscript]{batygin19_rev}.  However, this would only provide a potential constraint on $a_{P9}$, as the large libration amplitudes in mean anomaly with respect to Planet Nine would make it impossible to deduce the planet's current position in its orbit.  For a more detailed discussion, we direct the reader to \citet{bailey18}.

\bibliographystyle{apj}
\newcommand{\sci}{$Science$ }
\bibliography{n_1.bib}
\end{document}